\documentclass[twoside]{pltptentat}

\usepackage{natbib}
\usepackage{graphicx}
\usepackage{amssymb}

\bibpunct[, ]{(}{)}{,}{a}{}{,}
\begin{document}

\begin{frontmatter}



\title{QUANTIZED VORTICES IN ATOMIC BOSE-EINSTEN CONDENSATES}


\author{KENICHI KASAMATSU}
\address{Department of General Education, Ishikawa National College of Technology, Tsubata, Ishikawa 929-0392, Japan}
\author{MAKOTO TSUBOTA}
\address{Department of Physics, Osaka City University, Sugimoto 3-3-138, Osaka, Japan}

\begin{abstract}
In this review, we give an overview of the experimental and theoretical advances in the physics of quantized vortices in dilute atomic-gas Bose--Einstein condensates in a trapping potential, especially focusing on experimental research activities and their theoretical interpretations. Making good use of the atom optical technique, the experiments have revealed many novel structural and dynamic properties of quantized vortices by directly visualizing vortex cores from an image of the density profiles. These results lead to a deep understanding of superfluid hydrodynamics of such systems. Typically, vortices are stabilized by a rotating potential created by a laser beam, magnetic field, and thermal gas. Finite size effects and inhomogeneity of the system, originating from the confinement by the trapping potential, yield unique vortex dynamics coupled with the collective excitations of the condensate. Measuring the frequencies of the collective modes is an accurate tool for clarifying the character of the vortex state. The topics included in this review are the mechanism of vortex formation, equilibrium properties, and dynamics of a single vortex and those of a vortex lattice in a rapidly rotating condensate.
\end{abstract}

\end{frontmatter}

\newpage
\thispagestyle{empty}
\tableofcontents
\newpage

% main text
\section{Introduction} \label{intro}
The achievement of Bose--Einstein condensation in trapped atomic gases at ultra-low temperatures has stimulated intense experimental and theoretical activity in modern physics, as seen by the award of the Nobel Prize in Physics in 2001 \citep{Novel1,Novel2}. The Bose--Einstein condensate (BEC), a state of matter predicted  by Einstein in 1925, is created by the condensation of a macroscopically large number of bosons into one of the eigenstates of the single-particle density matrix below the Bose--Einstein transition temperature. A remarkable consequence of the condensation is an extension of microscopic quantum phenomena into the macroscopic scale. This is an essential origin of superfluidity and superconductivity, in which macroscopically extended phase coherence allows a dissipationless current to flow.

Superfluidity is closely related to the existence of quantized vortices. For weakly interacting BECs the superfluid velocity {\bf v} is given by the gradient of the phase $\theta$ of a ``condensate wave function'' ${\bf v}= (\hbar/m) \nabla \theta$ with the Planck constant $\hbar=h/2\pi$ and particle mass $m$. Since the wave function remains single-valued, the change in the phase around a closed contour must be an integer multiple of $2\pi$. Thus, the circulation $\Gamma$ around a closed contour is given by $\Gamma = \oint {\bf v} \cdot d {\bf l} = (h/m) q$ ($q = 0, 1, 2, \cdots $), which shows that circulation of a vortex is ``quantized" in units of $h/m$. Realization of weakly-interacting atomic-gas BECs has provided an ideal testing ground to study the physics of quantized vortices; up to now, several experimental groups have reported many interesting results. This experimental work has been followed by considerable theoretical activity, leading to proposals and new problems to be tackled. (For a review of the early research stages of quantized vortices, see \citep{Fetterrev}.)

In this article, we review the physics of quantized vortices in atomic-gas BECs, especially focusing on the progress of the experimental research. We restrict ourselves to arguments on {\it trapped} condensates with inhomogeneous density profiles. The aim of this review is to stimulate further developments of this field. By reflecting on the history of the current research and on unresolved problems, we hope to encourage researchers in low temperature physics to investigate quantized vortices in this system. While quantized vortices have been extensively studied in the field of superfluid helium \citep{Doneley}, there has been a resurgence of interest in vortices in atomic BECs because of the following reasons. First, the diluteness of a gas yields a relatively large healing length that characterizes the vortex core size, thus enabling the visualization of vortex cores by imaging techniques characteristic of this system (see Sec.\ref{imagingtech}). Because of this observational capability, the ability to manipulate a condensate wave function, and the tunability of the rotation over a wide range, these systems provide a unique approach to studying quantized vortices and their dynamics. Second, the finite size effect due to the trapping potential causes novel properties of vortices. Finally, multicomponent BECs provide new possibilities for studying unconventional vortex states that have been studied in other fields of physics, such as superfluid $^{3}$He, anisotropic superconductors, and theories in high-energy physics and cosmology \citep{Volovik}.

Since this is the first time that the topic of atomic BECs has appeared in the Progress of Low Temperature Physics, we start with a basic introduction to ultra-cold atomic systems, including how a condensate is formed and how they are manipulated. Although it is desirable to refer to the experimental and theoretical studies of such systems in detail, we will only mention the basic ideas necessary to understand some important issues in experiment and theory because of space restrictions. These issues are described in Sec. \ref{Basic}. More detailed accounts can be found in the comprehensive text books by Pethick and Smith \citep{Pethick}, Pitaevskii and S. Stringari \citep{Pitaevskii}, and in the review paper by Leggett \citep{Leggett}. In Sec. \ref{Vortexbasic}, we review the basic theory and experiments on quantized vortices in atomic BECs, addressing how vortices are created in this system and how they are detected. We also show the intrinsic mechanism of vortex nuclation and lattice formation in a trapped BEC. There are two interesting regimes classified by the rotation rate of the system: one at a slow rotation rate close to a critical rotation frequency where there is a single vortex, and another at high rotation rates for which a lattice of a large number of vortices is formed. The details of these two regimes are discussed in Sec. \ref{Singlevortex} and Sec. \ref{Vortexlattice}, respectively. Further interesting topics that cannot be explained sufficiently in this review and remaining future problems are presented in Sec. \ref{other}. We devote Sec. \ref{conclusions} to conclusions and outlook.

\section{Introduction to ultra-cold atomic-gas BECs} \label{Basic}
While progress toward the achievement of Bose--Einstein condensation in a dilute atomic gas had proceeded for the past few decades, researchers succeeded in creating a condensate in 1995 \citep{Anderson}. This realization brought great sensation in modern physics and opened a new research field combining condensed matter physics and atomic, molecular, optical (AMO) physics. Here, we briefly summarize the basic introduction about a system of ultra-cold atomic-gas BECs, giving background information for understanding the experiments and theories of quantized vortices in the following sections.

\subsection{General information}
A typical system considered here is a collection of neutral atoms with particle number $N \sim 10^{4}$ to $10^{7}$, trapped by a potential created by a magnetic field or an optical laser field. The density of the atomic gas is of the order of $n \simeq 10^{14}$ cm$^{-3}$, which is lower than that of air on the earth ($\sim 10^{19}$ cm$^{-3}$). The transition temperature to Bose-Einstein condensation can be estimated from a dimensional analysis of the relevant physical quantities ($m$, $n$, $\hbar$) as $k T_{c} \simeq \hbar^{2} n^{2/3}/m$, which is in a range from $100$ nK to a few $\mu$K. At such low temperatures, the gas phase cannot be a stable thermodynamic state and could in principle collapse to the solid phase. However, this relaxation is dominated by a three-body recombination process, which is a rare event for dilute and cold gases; the lifetime of the sample is thus long enough (of the order of a few seconds to a few minutes) to carry out experiments.

In typical experiments, there are several steps toward the condensation of atoms. The first is laser cooling, achieved with three pairs of counter-propagating laser beams along three orthogonal axes. Subsequently, the precooled gas is confined in a trapping potential, described typically by a harmonic potential. In this stage, the temperature is of order 100 $\mu$K, with 10$^{9}$ atoms. In the case of a trap  created by a magnetic field, the atoms are trapped by the Zeeman interaction of the electron spin with an inhomogeneous magnetic field. Thus, atoms with electron spins parallel to the magnetic field are attracted to the minimum of the magnetic field (weak-field seeking state), while ones with electron spin antiparallel are repelled (strong-field seeking state). Laser cooling alone cannot produce sufficiently high densities and low temperatures for condensation. The second step, evaporative cooling (a process in some sense similar to blowing on coffee to cool it), allows the removal of more energetic atoms, thus further cooling the cloud. The evaporation is effected by applying a radio-frequency magnetic field which flips the electron spin of the most energetic atoms. At the end of the process, the final temperature is about 100 nK and about 10$^{4}$-10$^{7}$ atoms remain.

Experimentally, the atomic-gas systems are attractive, since they can be manipulated by the use of lasers and magnetic fields. The cold gas is confined in a trap without microscopic roughness, as it is an extremely clean system. In addition, interactions between atoms may be affected either by using different atomic species or by changing the strength of an applied magnetic or electric field for species that exhibit Feshbach resonance. A further advantage is that, because of low density, the microscopic length scales are so large that the structure of the condensate wave function may be investigated by optical methods. Finally, the mean collision time $\tau_{\rm coll} \sim (\sigma n v)^{-1} \sim 10^{-3}$  sec between atoms ($\sigma$ is the cross section and $v$ the speed of atom) is comparable to the characteristic time of the collective mode, which prohibits a local equilibrium of the system. Thus, this system is ideal for studying nonequilibrium relaxation dynamics.

\subsection{Atomic Species}
The characteristics of BECs are mainly determined by atom--atom interactions, which depend crucially on the species of the condensed atoms. Most BEC experiments have been performed using alkali atoms because their ground state electronic structure is simple; all electrons except one occupy closed shells and the remaining electron is in an $s$ orbital in a higher shell. This structure is well suited to laser-based manipulation because its optical transitions can be excited by available lasers and the internal energy-level structure is favorable for cooling to very low temperatures. Since the first observation of BEC in atomic gases, BECs have been formed from nine different elements, including the alkali atoms $^{87}$Rb \citep{Anderson}, $^{23}$Na \citep{Davis}, $^{7}$Li \citep{Bradley}, H \citep{Fried}, $^{85}$Rb \citep{Cornish}, $^{41}$K \citep{Modugno}, $^{133}$Cs \citep{Weber}, and $^{39}$K \citep{Roati}, and the non-alkali atoms metastable He \citep{Robert,DosSantos}, $^{174}$Yb \citep{Takasu}, and $^{52}$Cr \citep{Griesmaier}. $^{87}$Rb and $^{23}$Na atoms are stable and have long lifetimes against inelastic collisional decay and are thus popular atomic species for BEC experiments.

\subsection{Detection}\label{imagingtech}
Once a BEC has been created in a harmonic trap, it is probed for its properties. This can be achieved either {\it in situ}, i.e., with the condensate inside the trap, or using a time-of-flight (TOF) technique. Although {\it in situ} diagnostics, such as nondestructive phase-contrast imaging \citep{Andrews}, are valuable tools for some applications, the TOF technique is more often used in vortex experiments. The TOF technique involves switching off the trapping field (magnetic or optical) at time $t=0$ and taking an image of the BEC a few (typically 5 to 25) milliseconds later. Switching off the trap allows the sample to expand before applying the laser beam probe, because the probe is difficult to apply at high densities. Images of the sample are most often taken by absorption, i.e., shining a resonant laser beam into the atomic cloud and using a CCD camera to observe the shadow cast by the absorption of photons, from which can be determined the integrated atomic density. This method is inherently destructive since real absorption processes are involved by spontaneous radiation and the accompanying heating.

\subsection{Manipulation}\label{manipu}
(i) {\it Laser created potential}

Atoms in a laser field experience a force, due mainly to the interaction of the laser field with the electric dipole moment induced in the atoms. The force on atoms in a laser field is used in a variety of ways in BEC experiments.

The character of the force is determined by the detuning given by $\Delta \equiv \hbar \omega_{\rm las} - (E_{e} - E_{g})$, where $\omega_{\rm las}$ is the laser frequency and $E_{g}$ ($E_{e}$) the ground (excited) state energy of an atom. The force also depends on the laser-beam intensity $I_{0}$, given by $I_{0} = \epsilon_{0} c \Gamma^{2} / d^{2}$, where $\epsilon_{0}$ is the dielectric constant, $c$ the speed of light, $d$ an appropriately defined dipole matrix element for the transition in question, and $\Gamma \equiv \hbar/\tau_{e}$ with the lifetime of the excited state $\tau_{e}$. In the limit $\Gamma \ll \Delta$, the change in energy of the atom in the laser field is $\Delta E_{\rm laser}({\bf r}) = (I({\bf r})/I_{0}) \Gamma^{2}/\Delta$. A region of high laser intensity thus provides an attractive potential for $\Delta < 0$ (``red detuning'') and a repulsive potential for $\Delta > 0$ (``blue detuning''). A red-detuned potential has been used as an optical trap for atoms. A blue-detuned potential creates a potential barrier that separates a condensate and an impurity (obstacle) potential. The interference pattern created by counter-propagating laser beams yields a periodic potential for atoms, called an optical lattice.
\\

(ii) {\it Hyperfine state}

Atomic BECs can have internal degrees of freedom, attributed to the hyperfine spin of atoms. A hyperfine-Zeeman sublevel of an atom with total electronic angular momentum ${\bf J}$ and nuclear spin ${\bf I}$ may be labeled by the projection $m_{F}$ of total atomic spin ${\bf F} = {\bf I} + {\bf J}$ on the axis of the field ${\bf B}$ and by the value of total $F$, which can take a value from $|I-J|$ to $|I+J|$. This is because the hyperfine coupling, which is proportional to ${\bf I} \cdot {\bf J}$, is much larger than the typical temperature of an ultra-cold atomic system. The hyperfine state is denoted by $| F, m_{F} \rangle$ with $m_{F} = -F, -F+1, \cdots , F-1, F$. The simultaneous trapping of atoms with different hyperfine sublevels makes it possible to create multicomponent (often called ``spinor'') BECs with internal degrees of freedom \citep{Ho1,Ohmi}, characterized by multiple order parameters \citep{Hall,Stenger,Barrett,Schmaljohann,Chang,Kuwamoto}.

An external field can couple the internal sublevels of the atom and cause coherent transition of the population. This coherent transition can be used to control the spatial variation of the condensate wave functions, resulting in an ``imprinting'' of a phase pattern onto the condensate. In most schemes, the spatial configuration of the field, the intensity and detuning of the laser fields, and the phase relationship between the different fields need to be carefully controlled to create the right phase pattern, that takes full advantage of the complex internal dynamics.
\\

(iii) {\it Feshbach resonance}

A salient feature of cold atom systems is that field-induced Feshbach resonance can tune the scattering length between atoms \citep{Inouye}, which determines the atom--atom interaction. A Feshbach resonance occurs when a quasi-bound molecular state in a closed channel has energy equal to that of two colliding atoms in an open channel. Such resonances can greatly effect elastic and inelastic collisions such as dipolar relaxation and three-body recombination.

Scattering near the resonance can be quantified by perturbation theory. To first order in the coupling between open and closed channels, the scattering is unaltered, because there are no continuum states in the closed channels. However, two particles in an open channel can scatter to an intermediate state in a closed channel, which subsequently decays to give two particles in an open channel. Considering such second-order processes, we can obtain the contribution to the scattering length as $\sim (E_{\rm op}-E_{\rm cl})^{-1}$,
where $E_{\rm op}$ is the energy of the particles in the open channel and $E_{\rm cl}$ is the energy of a state in the closed channels. Consequently, there are large effects if the energy $E_{\rm op}$ of the two particles in the entrance channel is close to the energy $E_{\rm cl}$ of a bound state in the closed channels. Therefore, coupling between the channels yields a repulsive interaction if the energy of the scattering particles is greater than that of the bound state, and an attractive interaction if it is less. Since the energies of the states depend on external parameters such as the magnetic field, the resonances can be used to control the interaction between atoms.

\subsection{Basic theory of trapped BECs}
\subsubsection{The Gross--Pitaevskii equation}
From a theoretical point of view, a major advantage of weakly-interacting atomic-gas BECs is that almost all the atoms in the system occupy the same quantum state and the condensate may be described very well in terms of mean-field theory. This is in contrast to liquid $^{4}$He, for which a mean-field approach is inapplicable due to the strong  correlations induced by the interactions between the atoms. Bogoliubov's treatment of a uniform Bose gas at zero temperature provides a useful mean-field description of a condensate. Subsequently, Gross and Pitaevskii independently considered an inhomogeneous dilute Bose gas, generalizing Bogoliubov's approach to include nonuniform states, which includes quantized vortices. Such nonuniform states of a dilute Bose gas can be understood by considering the second-quantized many-body Hamiltonian
\begin{eqnarray}
\hat H = \int d {\bf r} \hat{\Psi}^{\dagger}({\bf r})  \left[ - \frac{\hbar^{2} \nabla^{2} }{2m} + V_{\rm ex} ({\bf r}) + \frac{1}{2}  \int d {\bf r}' \hat{\Psi}^{\dagger}({\bf r}') V_{\rm int}({\bf r} - {\bf r}')  \hat{\Psi}({\bf r}')  \right] \hat{\Psi}({\bf r}), \label{secondhamiltonian}
\end{eqnarray}
expressed in terms of Boson field operators $\hat{\Psi}({\bf r})$ and $\hat{\Psi}^{\dagger }({\bf r})$ that obey Bose--Einstein commutation relations $[\hat{\Psi}({\bf r}),\hat{\Psi}^{\dagger}({\bf r}')]= \delta ({\bf r-r}')$, $\quad [\hat{\Psi}({\bf r}),\hat{\Psi}({\bf r}')] = [\hat{\Psi}^{\dagger}({\bf r}), \hat{\Psi}^{\dagger}({\bf r}')] = 0$. Here, $V_{\rm ex}({\bf r})$ is a trapping potential. The interparticle potential $V_{\rm int}$ is approximated by a short-range interaction $V_{\rm int} \simeq g \delta ({\bf r-r}')$, where $g = 4\pi \hbar^{2} a / m $ is a coupling constant, characterized by the s-wave scattering length $a$, because only binary collisions at low energy are relevant in a dilute cold gas and these collisions are independent of the details of the two-body potential.

In three dimentions, a remarkable feature of a dilute Bose gas at zero temperature is the existence of a macroscopic wave function $\Psi$ (an ``order parameter''). The macroscopic occupation of condensed particles makes it natural to write the field operator as a sum $\hat{\Psi}({\bf r},t) = \Psi({\bf r},t) + \hat{\phi}({\bf r},t)$ of a classical field $\Psi({\bf r},t)$ that characterizes the condensate and a quantum field $\hat{\phi}({\bf r},t)$ representing the remaining noncondensed particles. In order to derive the equation of motion for the order parameter, we write the time evolution of the operator $\hat{\Psi}({\bf r},t) = \exp(i \hat{H} t / \hbar) \hat{\Psi}({\bf r}) \exp(-i \hat{H} t / \hbar) $ using the Heisenberg equation with the many-body Hamiltonian
\begin{equation}
i \hbar \frac{\partial \hat{\Psi}({\bf r},t)}{\partial t} = [\hat{\Psi}({\bf r} ,t), \hat{H}] =  \left[- \frac{\hbar^{2} \nabla^{2}}{2m} + V_{\rm ex} + g \hat{\Psi}^{\dagger}({\bf r},t) \hat{\Psi}({\bf r},t) \right] \hat{\Psi}({\bf r},t). \label{operatoreq}
\end{equation}
To leading order, the Bogoliubov approximation neglects the noncondensed contribution, giving the time-dependent Gross--Pitaevskii (GP) equation
\begin{equation}
i \hbar \frac{\partial \Psi({\bf r},t)}{\partial t} = \left[ - \frac{\hbar^{2} \nabla^{2}}{2m} +V_{\rm ex} + g |\Psi({\bf r},t)|^{2} \right] \Psi({\bf r},t) \label{GPeq}
\end{equation}
for the condensate wave function $\Psi({\bf r},t)$. The GP equation (\ref{GPeq}) can be used to explore the dynamic behavior of the system, characterized by variations of the order parameter over distances larger than the mean distance between atoms. This equation is valid when the s-wave scattering length is much smaller than the average distance between atoms, and the number of atoms in the condensate is much larger than unity.

The ground state of a trapped BEC can be expressed within the formalism of the GP theory. We can write the condensate wave function as $\Psi({\bf r},t)=\Phi({\bf r}) e^{-i \mu t/\hbar}$, where $\Phi({\bf r})$ obeys the time-independent GP equation
\begin{equation}
\left[ - \frac{\hbar^{2} \nabla^{2}}{2m} +V_{\rm ex} + g |\Phi({\bf r})|^{2} \right] \Phi({\bf r}) = \mu \Phi({\bf r});  \label{staGPeq}
\end{equation}
$\Phi$ is normalized to the number of condensed particles $\int d {\bf r} |\Phi({\bf r})|^{2} = N_{0}$, which determines the chemical potenteial $\mu$. Typically, studies of trapped atomic gases involve the dilute limit (the gas parameter $\bar{n} |a|^3$ is typically less than 10$^{-3}$, where $\bar{n}$ is the average density of the gas), so that depletion of the condensate is small with $N^{\prime }=N-N_0\propto \sqrt{\bar n|a|^3}N\ll N$. Hence, most of the particles remain in the condensate such that $N_0 \simeq N$. The time-independent GP equation (\ref{staGPeq}) is also derived by minimizing the GP energy functional
\begin{equation}
E[\Psi, \Psi^{\ast}] = \int d {\bf r} \Psi^{\ast} \left( -\frac{\hbar^{2} \nabla^{2}}{2m}  + V_{\rm ex} + \frac{g}{2} |\Psi|^{2} \right) \Psi \equiv E_{\rm kin} + E_{\rm tr} +E_{\rm int}, \label{Hamilfunctional}
\end{equation}
subject to the constraint of a fixed particle number $N$. This constraint is taken into account by the Lagrange multiplier method; we write the minimization procedure as $\delta (E - \mu N)/ \delta \Psi^{\ast} = 0$, where the chemical potential $\mu$ is the Lagrange multiplier that ensures a fixed $N$.

Equation (\ref{staGPeq}) provides a starting point for studying the structure of a condensate in a harmonic confining potential $V_{\rm ex} = m ( \omega_{x}^2 x^2 + \omega_{y}^2 y^2 + \omega_{z}^2 z^2) / 2$. This introduces the length scale $a_{\rm ho} = \sqrt{\hbar / m \omega}$ with  $\omega = (\omega_{x} \omega_{y} \omega_{z})^{1/3}$. Although  the exact ground state can be obtained only by solving Eq. (\ref{staGPeq}) numerically, an approximate analytic solution can be gained when the interaction energy $E_{\rm int}$ is much larger than $E_{\rm kin}$ \citep{Baym}. To see this argument, let us neglect the anisotropy of the harmonic potential and assume that the cloud occupies a region of radius $\sim R$, so that $n \sim N/R^{3}$. Then, the scale of the harmonic oscillator energy per particle is $\sim m \omega^{2} R^{2} /2 $ while each particle experiences an interaction with the other particles of energy $\sim gN/R^{3}$. By comparing these energies, the radius is found to be $R \sim a_{\rm ho} (8 \pi N a/a_{\rm ho})^{1/5}$. The kinetic energy is of order $\hbar^{2}/2mR^{2}$, so that the ratio of the kinetic to interaction (or trap) energies is $\sim (N a/a_{\rm ho})^{-4/5}$. In the limit $Na/a_{\rm ho} \gg 1$, which is relevant to current experiments on trapped BECs, the repulsive interactions significantly expand the condensate, so that the kinetic energy associated with the density variation becomes negligible compared to the trap and interaction energies. As a result, the kinetic-energy operator can be omitted in Eq. (\ref{staGPeq}), giving the Thomas--Fermi (TF) parabolic profile for the ground-state density
\begin{equation}
n({\bf r}) \simeq |\Psi_{\rm TF}({\bf r})|^2 = \frac{\mu - V_{\rm ex}({\bf r})}{g} \Theta\left[ \mu - V_{\rm ex}({\bf r}) \right], \label{TFparabola}
\end{equation}
where $\Theta(x)$ denotes the step function. The resulting ellipsoidal density in three-dimensional (3D) space is characterized by two types of parameters: the central density $n_{0} = \mu/g$ and the three condensate radii $R_{j}^2 = 2\mu/m\omega_{j}^{2}$ ($j=x,y,z$). The chemical potential $\mu$ is determined by the normalization $\int d {\bf r} n({\bf r}) = N$ as $\mu = (\hbar \omega/2) (15 N a/a_{\rm ho})^{2/5} $.

\subsubsection{The Bogoliubov--de Gennes equation}
The spectrum of elementary excitations of a condensate is an essential ingredient  in calculations of the thermodynamic properties. To study the low-lying collective-excitation spectrum of trapped  BECs, the Bogoliubov--de Gennes (BdG) equation coupled with the GP equation is a useful formalism.

Let us consider the equation of motion for a small perturbation around the stationary state $\Phi$, which is a solution of Eq. (\ref{staGPeq}). The wave function takes the form $\Psi({\bf r},t) = [\Phi({\bf r}) + u_{j} ({\bf r}) e^{-i \omega_{j} t} - v_{j}^{\ast} ({\bf r}) e^{i \omega_{j} t}] e^{-i \mu t}$. By inserting this ansatz into Eq. (\ref{GPeq}) and retaining terms up to first order in $u$ and $v$, we obtain the BdG equation:
\begin{eqnarray}
\left(
\begin{array}{cc}
{\cal L}({\bf r}) & - g \Phi({\bf r})^{2}  \\
g \Phi^{\ast}({\bf r})^{2} & -{\cal L}({\bf r})
\end{array}
\right)  \left(
\begin{array}{c}
u_{j}({\bf r}) \\
v_{j}({\bf r})
\end{array}
\right) = \hbar \omega_{j} \left(
\begin{array}{c}
u_{j}({\bf r}) \\
v_{j}({\bf r})
\end{array} \right), \label{BdG}
\end{eqnarray}
where ${\cal L} ({\bf r})= - \hbar^{2} \nabla^{2}/2m + V_{\rm ex} - \mu + 2g |\Phi({\bf r})|^{2}$, and $\omega_{j}$ are the eigenfrequencies related to the quasiparticle normal-mode functions $u_{j}({\bf r})$ and $v_{j}({\bf r})$. The mode functions are subject to the orthogonality and symmetry relations $\int d {\bf r} [u_{i} u_{j}^{\ast} - v_{i} v_{j}^{\ast}] = \delta_{ij}$ and $\int d {\bf r} [u_{i} v_{j}^{\ast} - v_{i} u_{j}^{\ast}] = 0$.

Since the energy $\hbar \omega_{j}$ of this quasiparticle is defined with respect to the condensate energy, in Eq. (\ref{Hamilfunctional}) with the stationary solution $\Phi$, the presence of quasiparticles with negative frequencies implies an energetic (thermodynamic) instability for the solution $\Phi$. If there is energy dissipation in the system, the excitation of negative-energy modes lowers the total energy of the system and $\Phi$ relaxes to a more stable solution. We note that, since the matrix element of Eq. (\ref{BdG}) is non-hermitian, the eigenfrequencies can be complex-valued. When there are complex-valued frequencies, small-amplitude fluctuations of the corresponding eigenmodes grow exponentially during the energy-conserving time development. This is known as dynamical instability and is a main origin of the generation of nonlinear excitations, such as vortices or solitons, in ultracold condensates.

\section{Vortex formation in atomic BECs} \label{Vortexbasic}
In this section, we describe some basic properties of quantized vortices in trapped BECs and experimental procedures to create them. We also describe the nucleation mechanisms of quantized vortices in trapped BECs, including controlled schemes of phase engineering. Typically, experiments have used a smooth rotating potential, which is created by a laser or magnetic field, with a small transverse anisotropy to rotate the condensate. This potential induces a low-energy collective oscillation or shape deformation of the condensate. Such global motions of the condensate are responsible for the instability of the vortex nucleation, producing interesting nonequilibrium dynamics in the system. This is contrary to superfluid helium systems, where vortex nucleation occurs locally through roughness or impurities in the rotating container.

\subsection{Theoretical background}\label{vortexbasictheory}
As a simple example, let us consider the structure of a single vortex in a condensate trapped by an axisymmetric harmonic potential $V_{\rm ex}(r,z)=m \omega_{\perp}^{2} (r^{2}+\lambda^{2}z^{2})/2$ with aspect ratio $\lambda=\omega_{z}/\omega_{\perp}$. The condensate wave function with a quantized vortex line located along the $z$-axis takes the form $\Phi({\bf r}) = \phi(r,z) e^{iq\theta}$ with winding number $q$ and cylindrical coordinate $(r,\theta,z)$. $\phi$ is a real function related to the condensate density as $n (r,z)= \phi^{2}$. The velocity field around the vortex line is ${\bf v}_{\rm s} = (q \hbar/m r) \hat{\theta}$. Equation (\ref{staGPeq}) becomes
\begin{equation}
\left[ -\frac{\hbar^{2}}{2m} \left( \frac{\partial^{2}}{\partial r^{2}} + \frac{1}{r} \frac{\partial}{\partial r} + \frac{\partial^{2}}{\partial z^{2}} \right) + \frac{q^{2} \hbar^{2}}{2mr^{2}} + V_{\rm ex}(r,z) +g n \right] \phi
= \mu \phi. \label{singletimeindGPrz}
\end{equation}
Here, the centrifugal term $q^{2} \hbar^{2}/2mr^{2}$ arises from the azimuthal motion of the condensate. Equation (\ref{singletimeindGPrz}), solved numerically, gives the structure of the vortex. In the TF limit $N a/a_{\rm ho} \gg 1$, we can omit the terms involving derivatives with respect to $r$ and $z$ in Eq. (\ref{singletimeindGPrz}), and the density can be obtained approximately as
\begin{eqnarray}
n (r,z)  = n_{0} \left( 1 - \frac{r^{2}}{R_{\perp}^{2}} - \frac{z^{2}}{R_{z}^{2}} - q^{2} \frac{\xi^{2}}{r^{2}} \right) \Theta \left( 1 - \frac{r^{2}}{R_{\perp}^{2}} - \frac{z^{2}}{R_{z}^{2}} - q^{2} \frac{\xi^{2}}{r^{2}}  \right), \label{Thoamvor}
\end{eqnarray}
where $n_{0}=\mu/g$ is the density at the center of the {\it vortex-free} TF profile. We define the TF radius $R_{\perp}^{2} = 2 \mu / m \omega_{\perp}^{2}$ and $R_{z}^{2} = 2 \mu / m \omega_{z}^{2}$ and the healing length $\xi = (\hbar^{2}/2 m g n_{0})^{1/2} = (8 \pi a n_{0}) ^{-1/2}$. Equation (\ref{Thoamvor}) shows that the condensate density vanishes at the center, out to a distance of order $\xi$, due to the centrifugal term $\xi^{2}/r^{2}$ (numerical solution shows that the density grows as $r^{2}$ away from the center), whereas the density in the outer region has the form of an upward-oriented parabola. Hence, the healing length $\xi$ characterizes the vortex core size; for typical BEC parameters, $\xi \sim 0.2$ $\mu$m. In the TF limit, the core size is very small because $\xi/R_{\perp} = \hbar \omega_{\perp} /2 \mu = (15 Na/a_{\rm ho})^{-2/5} \ll 1$. Increasing the winding number $q$ widens the core radius due to the centrifugal effects.

The energy associated with a single vortex line is an important quantity to determine the stability of the vortex state. The dominant contribution to this energy is the kinetic energy of the superfluid flow by a vortex. The energy is estimated as
\begin{equation}
E_{1} = \int \frac{1}{2} m n v_{\rm s}^{2} d {\bf r} \simeq \frac{m \bar{n}}{2} R_{z} \int^{R_{\perp}}_{\xi} v_{\rm s}^{2} 2 \pi r d r = q^{2} R_{z} \frac{\pi \hbar^{2} \bar{n}}{m} \ln \left( \frac{R_{\perp}}{\xi} \right), \label{onevorenrgy}
\end{equation}
where we assume the spatially uniform density $\bar{n}$ and the size along the $z$-axis $R_{z}$. Since $E_{1} \propto q^{2}$, vortices with $q>1$ are energetically unfavorable. For example, the energy cost to create one $q=2$ vortex is higher than that to create two $q=1$ vortices. Therefore, a stable quantized vortex usually has $q=1$, except for that in non-simply connected geometry, and we will mainly concentrate on the $q=1$ vortex in the following discussions. In atomic BECs, however, such a multiply quantized vortex can be created experimentally by using topological phase imprinting \citep{Shin} and exhibits interesting disintegration dynamics, as discussed in Sec. \ref{splittingbec}.

It is necessary to ensure the stability of a vortex in a trapped BECs against a non-vortex state to investigate its behavior. Imposing rotation on the system is a direct way to achieve stabilization. If the system is under rotation, it is convenient to consider the corresponding rotating frame; for a rotation frequency ${\bf \Omega}=\Omega \hat{\bf z}$, the integrand of the GP energy functional (\ref{Hamilfunctional}) acquires an additional term,
\begin{equation}
E' = \int d {\bf r} \Psi^{\ast} \left( -\frac{\hbar^{2} \nabla^{2}}{2m} + V_{\rm ex} + \frac{g}{2} |\Psi|^{2} - \Omega L_{z} \right) \Psi , \label{GPenergyrot}
\end{equation}
where $L_{z}=- i \hbar (x \partial_{y} - y \partial_{x})$. The corresponding GP equation becomes
\begin{equation}
i \hbar \frac{\partial \Psi({\bf r},t)}{\partial t} = \left[ - \frac{\hbar^{2} \nabla^{2}}{2m} +V_{\rm ex} + g |\Psi({\bf r},t)|^{2}  - \Omega L_{z} \right] \Psi({\bf r},t) . \label{GPeqrot}
\end{equation}
If there is a quantized vortex along the trap axis, $\langle L_{z} \rangle = N \hbar$, so that the corresponding energy of the system in the rotating frame is $E_{1}'=E_{1} - N \hbar \Omega$. The difference between $E_{1}'$ and the vortex-free energy $E_{0}'$ gives the favorable condition for a vortex to enter the condensate. Since $E_{0}'$ is equal to the energy $E_{0}$ in the laboratory frame, the difference is given by $\Delta E' = E_{1}'-E_{0}' = E_{1}-E_{0} - N \hbar \Omega$. Thus, the critical rotation frequency $\Omega_{c}$ for the existence of an energetically stable vortex line is given by $\Omega_{c} = (E_{1}-E_{0})/N \hbar$. Above the critical rotation frequency $\Omega_{c}$ the single vortex state is ensured to be {\it thermodynamically} stable.

To calculate $E_{1}$ more quantitatively, it is necessary to take into account the inhomogeneous effect of the condensate density \citep{Lundh}. In the TF limit, for a condensate in a cylindrical trap $\omega_{z}=0$ (an effective 2D condensate), the critical frequency is given by $\Omega_{c} = (2 \hbar/ m R_{\perp}^{2}) \ln (0.888 R_{\perp}/\xi)$. For an axisymmetric trap $V_{\rm ex}(r,z)$, the critical frequency is $\Omega_{c} = (5\hbar / 2 m R_{\perp}^{2}) \ln (0.671 R_{\perp}/\xi)$. $\Omega_{c}$ for a nonaxisymmetric trap is slightly modified by a small numerical factor, and has been discussed analytically \citep{Svidzinsky2} and numerically \citep{Feder2}.

When the rotation frequency is significantly higher than $\Omega_{c}$, further vortices will appear in the form of a triangular lattice, in analogy to the Abrikosov lattice of magnetic fluxes in type-II superconductors. The nature of the equilibrium state changes, first to a state with two vortices rotating around each other, then to three vortices in a triangle form, and subsequently to arrays of more vortices \citep{Butts,Feder4}. The detail of this state will be described in Sec. \ref{Vortexlattice}.

\subsection{Vortex formation in a stirred condensate}
Rotation effects of atomic BECs were first studied based on the knowledge of deformed atomic-nuclei systems. A slight rotation of a deformed trap excites so-called ``scissors modes'', which are closely related to the {\it irrotationality} $\nabla \times {\bf v}_s = 0$ of the superfluid hydrodynamics \citep{Odelin, Marago}. Although these experiments demonstrated the superfluidity of atomic BECs, more intuitive evidence can be gained by observing quantized vortices. However, since the first experimental realization of BECs, there have been technical hurdles to rotating a system. For atomic gases trapped by an impurity-free external potential, it was supposed that rotation of the potential could not transfer a sufficient angular momentum into the condensate. Thus, the first experimental detection of a vortex in an atomic BEC, by Matthews {\it et al.} \citep{Matthews}, was made by using a complicated phase imprinting technique proposed by Williams and Holland \citep{Williams}. Subsequently, vortices have been created by stirring a condensate mechanically with an ``optical spoon''; the first success of this ``rotating bucket'' experiment was reported by Madison {\it et al}. \citep{Madison}. In the following, we detail the rotating bucket experiments.

\subsubsection{Experimental scheme to rotate condensates}\label{stirringexp}
\begin{figure}[btp]
\begin{center}
\includegraphics[height=0.165\textheight]{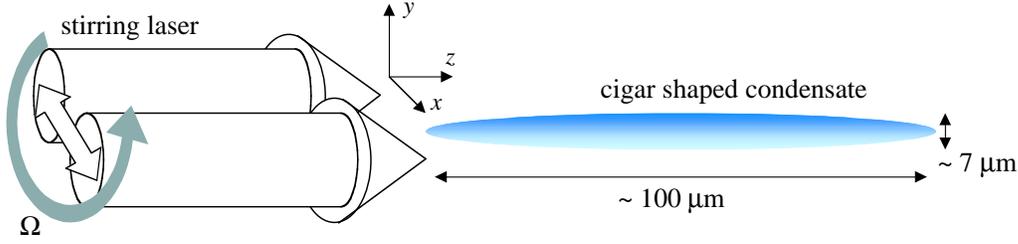}
\end{center}
\caption{Schematic illustration of the experimental setup used by the ENS group. A cigar-shaped condensate is rotated by an effective dipole potential made by the laser beam. The laser beam oscillates rapidly with a very large frequency around the $z$-axis, with an amplitude $\sim$ 16 $\mu$m. The beam width is about 20 $\mu$m. The dipole potential rotates with an angular frequency $\Omega$.}
\label{ENSschema}
\end{figure}
Madison {\it et al.} at Ecole Normale Sup\'{e}rieure (ENS) succeeded in observing a vortex and a vortex lattice in an atomic BEC \citep{Madison}, using a method similar to the observation of superfluid helium in a rotating bucket \citep{Yarmchuk}. A schematic illustration of their experimental setup is shown in Fig. \ref{ENSschema}. Since the magnetic trap is axially symmetric, its rotation cannot impart an angular momentum to the condensate. In this scheme, a laser beam is propagated along the $z$-axis of the condensate. This beam rapidly oscillates around the $z$-axis with a frequency much larger than the typical trapping frequency, which is effectively regarded to be as if two laser beams are located at an equilibrium position. Thus, the two laser beams break the axisymmetry of the condensate, allowing an angular-momentum transfer into the condensate. An optical spoon is then realized by rotating the two beams around the $z$-axis. Since the beam width is larger than the radial size of the condensate, the condensate is trapped in a trap combining an axisymmetric harmonic potential and a nonaxisymmetric harmonic potential created by a stirring laser beam. In a rotating frame, the combined potential can be written as $(1/2) m \omega_{\perp}^{2} [ (1 + \epsilon) X^{2} + (1-\epsilon) Y^{2} ] + (1/2) m \omega_{z}^{2} z^{2}$, where $X$ and $Y$ are the coordinates in the rotating frame, $\epsilon = (\omega_{X}^{2} - \omega_{Y}^{2}) / (\omega_{X}^{2} + \omega_{Y}^{2})$ is the anisotropic parameter, and $\omega_{\perp} = \sqrt{(\omega_{X}^{2} + \omega_{Y}^{2})/2}$. By rotating this potential at a frequency $\Omega$, a vortex is formed above a certain critical value of $\Omega$ after the equilibration. When $\Omega$ is increased further, multiple vortices appear, forming a triangular lattice. The quantized vortices can be directly visualized as ``dips'' in the transverse density profile of the TOF image.

\begin{figure}[hbtp]
\begin{center}
\includegraphics[height=0.158\textheight]{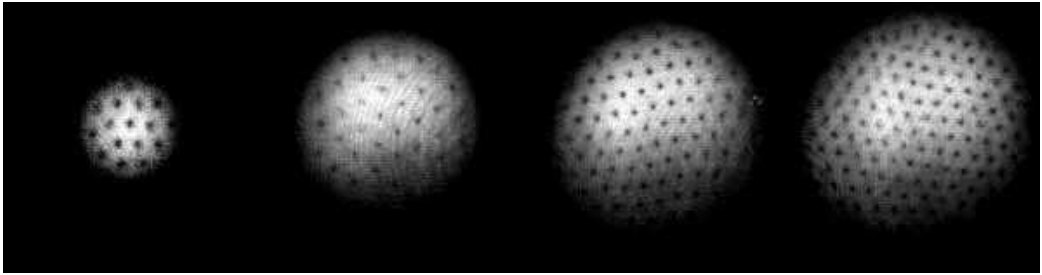}
\end{center}
\caption{Typical density profiles of a rotating condensate taken by TOF measurement. The condensates contain approximately 16, 32, 80, and 130 vortices from left to right. From J.R. Abo-Shaeer et al., SCIENCE 292, 476 (2001). Reprinted with permission from AAAS.}
\label{MITvordenang}
\end{figure}
Following the experiments of the ENS group, other groups have also observed quantized vortices using slightly different methods under the concept of the rotating bucket. Abo-Shaeer {\it et al.} at Massachusetts Institute of Technology (MIT) observed a vortex lattice consisting of up to 100 vortices in a $^{23}$Na condensate, as shown in Fig. \ref{MITvordenang}. $^{23}$Na condensates can be much larger than $^{87}$Rb condensates \citep{Abo}. Hodby {\it et al.} created a vortex lattice by rotating the anisotropic magnetic trap directly without using a laser beam, which is similar to the rotating bucket experiment \citep{Hodby}. This method has the advantage that a wider range of the anisotropic parameter $\epsilon$ can be selected than for the optical spoon. Rotating an optical spoon made by multiple-spot laser beams or narrow focusing beams has also used for nucleating vortices \citep{Raman2}.

In the above methods, the condensate was rotated by an external potential. Haljan {\it et al.} at Joint Institute for Laboratory Astrophysics (JILA), in contrast, created a vortex state by cooling an initially rotating thermal gas in a static confining potential \citep{Haljan}. Thermal gas above the transition temperature was rotated by a slightly anisotropic trap. After recovering the anisotropy of the potential, the rotating thermal gas was evaporatively cooled until most of the atoms were condensed. Although the atom number decreased through the evaporative cooling, the condensate continued to rotate because the angular momentum per atom did not change and thus the vortex lattice was created. This method allows the investigation of the intrinsic mechanism of vortex nucleation, which is independent of the character of the stirring potential. In addition, since atoms can be selectively removed during the evaporation, spinning up of the condensate can be efficiently achieved by removing atoms extending in the axial direction, as opposed to the radial direction, and hence a BEC with a high rotation rate can be obtained. 

\subsubsection{Theory of vortex nucleation and lattice formation}\label{formationdyn}
(i) {\it Surface mode instability}

The critical rotation frequency $\Omega_{c}$ indicates the energetic stability of the central vortex state and provides a lower bound for the critical frequency. Vortex nucleation of a non-rotating condensate occurs when the trap is rotated at a higher frequency than $\Omega_{c}$, to overcome the energy barrier that stops the transition from the nonvortex state to the vortex state \citep{Isoshima}. The threshold of the rotation frequency for instability, leading to vortex nucleation, is related to the excitation of surface modes of the trapped condensate \citep{Feder3,Dalfovo2,Anglin,Williams2,Simula2}. It has been shown that, according to the Landau criterion for rotationally invariant systems, the critical frequency is given by $\Omega_{v} = {\rm min}(\omega_{\ell} / \ell)$, where $\omega_\ell$ is the frequency of a surface mode with multipolarity $\ell$. Above $\Omega_{v}$, negative-energy surface modes appear with high multipolarities in the spectrum of a non-rotating condensate \citep{Isoshima,Dalfovo2}, which may lead to vortex generation. The negative-energy modes can grow only in the presence of dissipation, caused by e.g., thermal atoms \citep{Williams2}. This mechanism occurred in the experiment of the JILA group \citep{Haljan}. In this experiment, vortices were formed by cooling a rotating thermal cloud to below $T_c$, where the surface modes were excited by the ``wind'' of the rotating thermal cloud.

\begin{figure}[hbtp]
\begin{center}
\includegraphics[height=0.22\textheight]{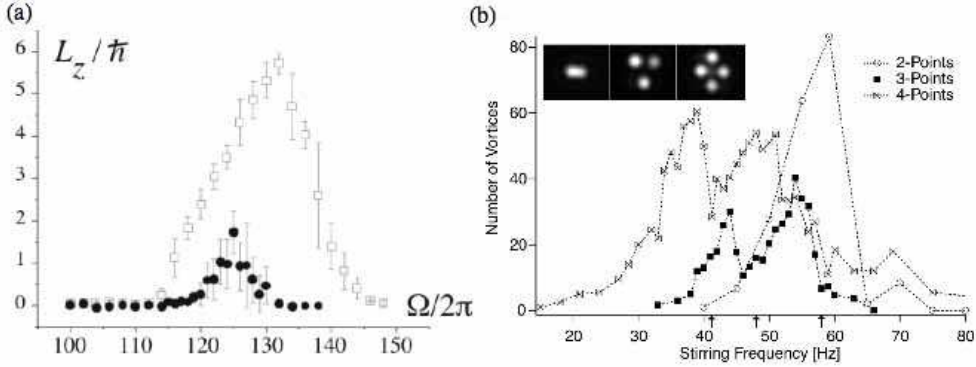}
\end{center}
\caption{Dependence of the vortex number (angular momentum per atom) on the  rotation frequency. (a) Result of the ENS group \citep{Chevy,Chevy2}. The parameter values are $N=2.5 \times 10^{5}$, $\omega_{\perp} = 2 \pi \times 172$ Hz, and an anisotropic parameter $\epsilon$ (black dots: $\epsilon=0.01$, white square: $\epsilon=0.02$). The angular momentum was measured by the surface-wave spectroscopic technique described in Sec. \ref{gyroscope}. With increasing $\Omega$, the value of the angular momentum has a peak at $\Omega \simeq \omega_{\perp}/\sqrt{2} = 122$ Hz, a resonance frequency of the quadrupole mode. The width of this peak structure increases as the trapping potential becomes more anisotropic. (b) Result of the MIT group \citep{Raman2}. The parameter values are $N \sim 10^{7}$ and $\omega_{\perp} = 2 \pi \times 86$ Hz. The arrows below the graph show the positions of the surface mode resonance $\omega_{\perp}/\sqrt{\ell}$. The inset shows 2-, 3-, and 4-point potentials produced by a laser beam. (Taken from \citep{Chevy2} and \citep{Raman2}. Reprinted with permission from American Physical Society (APS).)}
\label{ENSvordenang}
\end{figure}
The nucleation frequency $\Omega_{v}$ is insufficient to explain the results of the groups using external stirring potentials \citep{Madison,Abo,Hodby}. For example, in the case of the ENS group, the number of nucleated vortices has a peak near $\Omega = 0.7 \omega_{\perp}$, as shown in Fig. \ref{ENSvordenang}(a). These experiments confirm that instability occurs when a particular surface mode is resonantly excited by a deformed rotating potential. The optical spoon of the ENS group mainly excites the surface mode with $\ell =2$ (quadrupole mode). In a rotating frame with frequency $\Omega$, the frequency of the surface mode is increased by $-\ell \Omega$. This resonance thereby occurs close to the rotation frequency $\Omega = \omega_\ell /\ell$. In the TF limit, the dispersion relation for the surface mode is given by $\omega_{\ell}=\sqrt{\ell}\omega_{\perp}$ \citep{Stringari2}. Hence, it is expected that the quadrupole mode with $\ell =2$ is resonantly excited at $\Omega = \omega_{\perp}/ \sqrt{2} \simeq 0.707 \omega_{\perp}$. A theoretical study has revealed that, when the quadrupole mode is resonantly excited, an imaginary component appears in frequencies of fluctuations with high multipolarities \citep{Sinha}. This indicates that dynamic instability can trigger vortex nucleation even at zero temperature. This scheme is supported further by the experiment of the MIT group \citep{Raman2}, where surface modes with higher multipolarities ($\ell = 3, 4$) were resonantly excited using multiple laser-beam spots, where the largest number of vortices were generated at frequencies close to the expected values $\Omega = \omega_{\perp}/\sqrt{\ell}$, as seen in Fig. \ref{ENSvordenang}(b).

Another mechanism to produce vortex nucleation is formulated by considering elliptically-deformed stationary states of a BEC in a rotating elliptical trap. The stationary solutions can be obtained in the TF limit by solving the superfluid hydrodynamic equations \citep{Recati}. At low rotation rates, only one stationary solution is found (``normal branch"). However, at higher rotation rates, $\Omega > \omega_\perp/\sqrt{2}$ specifically, a bifurcation of the solutions occurs and up to three stationary solutions appear, with two solutions in an ``overcritical branch" and one normal-branch solution. Above $\Omega$ of the bifurcation point, one or more of the solutions become dynamically unstable for fluctuations with high multipolarities \citep{Sinha}, which leads to vortex formation \citep{Parker2}. Madison {\it et al.\ } followed these stationary states experimentally by adiabatically introducing trap ellipticity and rotation. They observed vortex nucleation in the expected dynamically unstable region \citep{Madison2}. 

An overview on the problem of vortex nucleation in a trapped BEC can be seen in the review paper \citep{Ghosh}.
\\

(ii) {\it Formation dynamics of a vortex lattice}

\begin{figure}[hbtp]
\begin{center}
\includegraphics[height=0.25\textheight]{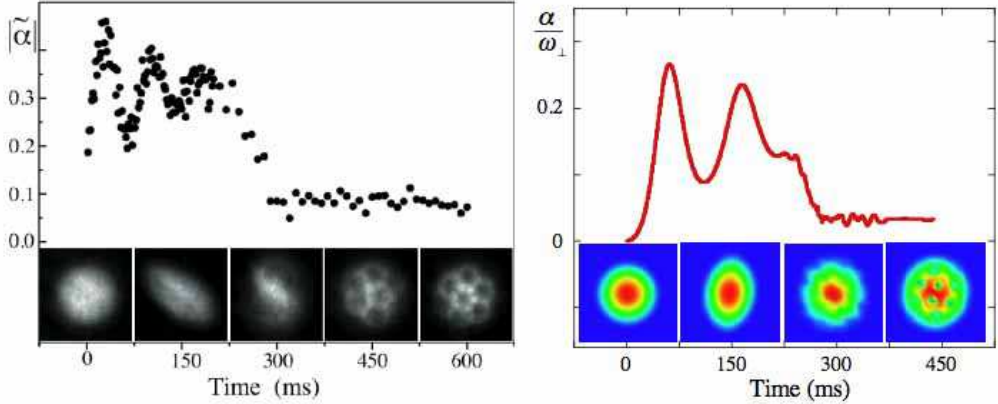}
\end{center}
\caption{(Left) Measurement of the time dependence of $\alpha $ (see text) when the stirring anisotropy is turned on rapidly from $\epsilon = 0$ to $\epsilon =0.025$ in 20 msec  and $\Omega=0.7$, and held constant for 300 ms. Five images taken at intervals of 150 ms show the transverse profile of the elliptic state and reveal the nucleation and ordering of the vortex lattice. (Taken from \citep{Madison2}. Reprinted with permission from APS.) (Right) Comparing numerical simulation of the 3D GP equation with phenomenological dissipation ($\gamma = 0.03$) and the experimentally relevant parameters. The density profile in the bottom panels is integrated along the $z$-axis.}
\label{ENSlattice}
\end{figure}
Madison {\it et al}. directly observed nonlinear processes such as vortex nucleation and lattice formation in a rotating condensate
\citep{Madison2}. The left panel of Fig. \ref{ENSlattice} depicts the time development of the condensate ellipticity $\alpha = \Omega (R_{X}^{2}-R_{Y}^{2})/(R_{X}^{2}+R_{Y}^{2})$, where $R_{X,Y}$ is the TF radius measured from the image. By suddenly turning on the rotation of the potential, the initially axisymmetric condensate undergoes a collective quadrupole oscillation in which the condensate deforms elliptically. This oscillation continues for a few hundred milliseconds with gradually decreasing amplitude. Then, the axial symmetry of the condensate suddenly recovers and concurrently the vortices enter the condensate from its surface, eventually settling into a lattice configuration.

This observation has been well reproduced by a simulation of the GP equation in 2D \citep{Tsubota,Kasamatsu} and 3D \citep{Kasamatsu2} space. The results shown in the right panel of Fig. \ref{ENSlattice} were obtained by a 3D simulation with the parameters of the ENS group experiment. The simulation result shows that after a few hundred milliseconds, the boundary surface of the condensate becomes unstable and generates ripples that propagate along the surface, identified as invisible ``ghost'' vortices in the low-density surface region. The ripples develop into vortex cores, which enter the condensate. In these simulations, a dissipation term was introduced phenomenologically by rewriting the time derivative term of the GP equation (\ref{GPeqrot}) as $(i-\gamma)\partial/\partial t$, which caused the lattice configuration to settle. Other works have simulated vortex lattice formation using dissipation derived from the microscopic approach, such as quantum kinetic theory \citep{Penckwitt} or the classical field formalism \citep{Lobo}. Long numerical propagation of the energy-conserving GP equation can cause crystallization of a lattice through the vortex--phonon interaction \citep{Parker}
\\

(iii) {\it Vortex nucleation by a moving object}

Vortices can also be nucleated in BECs by a moving localized potential. Numerical simulations of the GP equation for a 2D uniform condensate flow around a circular hard-walled potential show that vortex--antivortex pairs nucleate when the flow velocity exceed a critical value \citep{Frisch}. In trapped BECs, a similar situation can be realized experimentally using a narrow blue-detuned laser potential, being studied theoretically \citep{Jackson2,Crescimanno}. In experiments by the MIT group, a repulsive laser beam was oscillated in an elongated condensate to study the dissipationless flow of a Bose gas \citep{Raman,Onofrio}. Although vortices were not observed directly, the measurement of condensate heating and drag above a critical velocity was consistent with the nucleation of vortices \citep{Jackson3}. Focused laser beams moving in a circular path around the trap center can also stir the condensate by nucleating vortices \citep{Caradoc,Caradoc2,Lundh1}. This scheme was used in the experiment detailed in Ref. \citep{Raman2}, where vortices were generated at lower stirring frequencies than the critical value given by surface mode instability.

\subsection{Phase engineering}\label{phaseeng}
Atom optics techniques allow the controlled creation of vortices by imprinting a spatial phase pattern of the condensate wave function. Several ideas for the creation of vortices by this technique have been proposed theoretically, based on the coherent control of the time evolution of the wave function, instead of mechanical rotation \citep{Marzlin,Marzlin2,Dum,Williams,Nakahara,Andrelczyk,Damski,Nandi,Kapale,Mottonen2}.

\subsubsection{Phase-imprinting method}\label{imprintphase}
The first observation of a quantized vortex in an atomic-gas BEC was achieved in a two-component BEC consisting of $^{87}$Rb atoms with hyperfine spin states $|F=1,m_{F}=-1 \rangle \equiv |1 \rangle$ and $|F=2,m_{F}=1 \rangle \equiv |2 \rangle$ \citep{Matthews}, which were confined simultaneously in almost identical magnetic potentials. Since the scattering lengths between atoms of $ |1 \rangle$ and $|1 \rangle$, $ |2 \rangle$ and $|2 \rangle$, and $|1 \rangle$ and $|2 \rangle$ are all different, the two states are not equivalent, and the two-component condensate is characterized by two-component order parameters.

In this experiment, condensed atoms are initially trapped in one state, say, the $|1\rangle$ state. Then, a two-photon microwave field is applied, inducing coherent Rabi transitions of atomic populations between the $|1\rangle $ state and the $|2\rangle$ state. For a homogenous system in which both components have uniform phases, interconversion takes place at the same rate everywhere. However, the time variation of the spatially inhomogeneous potential changes the nature of the interconversion. This is a key point of the phase-imprinting method for vortex creation. 

The underlying physics can be understood by considering a co-rotating frame with an off-centered perturbation potential at the rotation frequency $\Omega'$. In this frame, the energy of a vortex with one unit of angular momentum is shifted by $\hbar \Omega'$ relative to its value in the laboratory frame. When this energy shift is compensated for by the sum of the detuning energy of an applied microwave field and the small chemical potential difference between the vortex and non-vortex states, a resonant transfer of population can occur. Experimentally, the rotating perturbation is created by a laser beam with a spatially inhomogeneous profile, rotating around the initial nonrotating component, say $| 1\rangle$. By adjusting the detuning and $\Omega'$, the $|2\rangle$ component is resonantly transferred to a state with unit angular momentum by precisely controlling the time when the coupling drive is turned off \citep{Williams}. This procedure results in a ``composite'' vortex, where the $|2\rangle$ component has a vortex at the center, whereas the nonrotating $|1\rangle$ component occupies the center and works as a {\it pinning} potential that stabilizes the vortex core \citep{Kasamatsurev}.

As shown in Sec. \ref{manipu}, a far-off-resonant laser beam can create an external potential $V_{\rm las}({\bf r})$ in the condensate. By directly applying a laser pulse with an inhomogeneous intensity to the condensate, the condensate phase can be modulated. This can be easily understood by observing the evolution of the phase by inserting the form $\Psi({\bf r},t) = |\Psi({\bf r},t) | \e^{i \theta({\bf r},t)}$ into Eq. (\ref{GPeq}). When the laser intensity is much stronger than the other terms and the duration of the pulse $\tau$ is sufficiently short, the evolution of the phase is governed by $\theta ({\bf r}, t) = - \hbar^{-1} \int_{0}^{\tau} dt V_{\rm las}({\bf r},t)$. Since the potential amplitude of $V_{\rm las}$ is proportional to the laser intensity, a suitable spatial variation of the intensity can imprint the phase into the condensate \citep{Andrelczyk}. This phase-imprinting technique has been used to create a {\it dark soliton} \citep{Burger,Denschlag}, which is a topological excitation with a complete density dip across which the phase changes by $\pi$. It is known that a dark soliton in a dimensional space larger than 2D experiences dynamical instability, called ``snake instability''. This instability causes the decay of the dark solitons into a form of a {\it vortex ring} \citep{Anderson2,Dutton}.

\subsubsection{Topological vortex formation}\label{topologicalphase}
Leanhardt {\it et al.} \citep{Leanhardt,Leanhardt2} used a method called ``topological phase imprinting'' \citep{Nakahara,Isoshima2,Ogawa} to create a vortex in a trapped BEC. In this experiment, $^{23}$Na condensates were prepared in either a lower, $|F,m_F\rangle = |1,-1\rangle$, or upper, $|2,+2\rangle$, hyperfine state and confined in a Ioffe--Pritchard magnetic trap, described by ${\bf B} = B' (x \hat{x} - y \hat{y}) + B_{z} \hat{z}$. A vortex was created by adiabatically inverting the axial bias field $B_{z}$ along the trap axis.

To interpret the mechanism, consider an alkali atom with a hyperfine-spin $|F|=1$. The order parameter has three components $\Psi_{\pm 1}$ and $\Psi_0$ corresponding to $F_z=\pm 1, 0$, respectively. The basis vectors in this representation are $\{|\pm \rangle, |0 \rangle\}$. We introduce another set of basis vectors $|x \rangle, |y \rangle$ and $|z \rangle$, which are defined by $F_x |x \rangle= F_y |y \rangle=F_z |z \rangle=0$. These vectors are related to the previous vectors as $|\pm 1\rangle = \mp (1/\sqrt{2})\left( |x \rangle \pm i|y \rangle\right)$ and $|0 \rangle=|z \rangle$. When the $z$-axis is taken parallel to the uniform magnetic field, the order parameter of the weak field seeking state takes the form $\Psi_{-1} = \psi$ and $\Psi_{0}=\Psi_{1}=0$, or in vectorial form as ${\bf \Psi} =(\psi/\sqrt{2}) \left(\hat{\bf x} -i \hat{\bf y} \right)$. When the magnetic field points in the direction $\hat{\bf B} =(\sin \beta \cos \alpha, \sin \beta \sin \alpha, \cos \beta)$, a rotational transformation with respect to the Euler angle $(\alpha, \beta, \gamma)$ gives ${\bf \Psi} = (\psi/\sqrt{2}) e^{i \gamma} (\hat{\bf m} + i \hat{\bf n})$, where $\hat{\bf m} = (\cos \beta \cos \alpha, \cos \beta \sin \alpha, -\sin \beta)$ and $\hat{\bf n}= (\sin \alpha, - \cos \alpha, 0)$. The unit vector $\hat{\bf l} = \hat{\bf m} \times \hat{\bf n} = (\cos \alpha \sin \beta, - \sin \alpha \sin \beta, - \cos \beta)$ is the direction of the spin polarization. The three real vectors $\{\hat{\bf l}, \hat{\bf m}, \hat{\bf n}\}$ form a triad, analogous to the order parameter of the orbital part of superfluid $^{3}$He. The same amplitudes in the basis $\{|0 \rangle, |\pm \rangle\}$ are $\Psi_{1} = (\psi/2) (1 - \cos \beta) e^{-i \alpha + i \gamma} $, $\Psi_0 = - (\psi/\sqrt{2}) \sin \beta e^{i \gamma}$, and $\Psi_{-1} = (\psi/2) (1 + \cos \beta) e^{i \alpha + i \gamma}$.

When the field $B_z$ is strong compared to the quadrupole field, the trapped condensate has an order parameter ${\bf \Psi} = (\psi/\sqrt{2})(\hat{\bf x} - i \hat{\bf y})$ without vorticity. This configuration corresponds to $\beta=0$ and $\gamma=-\alpha=\phi$, where $\phi$ is the azimuthal angle. Then, $B_z$ is adiabatically decreased, where the adiabatic condition is required for atoms to remain in the weak field seeking state so that $\hat{\bf l}$ is always antiparallel to ${\bf B}$. In the final step, the external field $B_{z}$ is gradually increased in the opposite ($-z$) direction. Then, the $\hat{\bf l}$-vector points so that $\beta= \pi$. Substituting these angles into $\Psi_{\pm 1}$ and $\Psi_{0}$, we obtain $\Psi_{-1} = \Psi_0 = 0$ and $\Psi_{1}  =  \psi e^{2i \phi}$, which corresponds to a vortex with winding number $q=2$. This result can be reinterpreted in terms of Berry's phase \citep{Ogawa,Leanhardt} (which is why it is called ``topological phase imprinting''). When the hyperfine spin is $F$ in general, we obtain a vortex with a winding number $2F$ since $\Psi_{-F}$ and $\Psi_{F}$ have phases $F(\alpha+\gamma)$ and $F(-\alpha+\gamma)$, respectively \citep{Shin,Kumakura}.

When the bias field vanishes during the inversion ($B_{z} = 0$), a spin texture known as cross disgyration appears in the Ioffe--Pritchard trap. Here, the angle $\beta$ increases from $0$ to $\pi/2$ and $\gamma$ and $-\alpha$ are identified with $\phi$, where the $\hat{\bf l}$-vector aligns with a hyperbolic distribution around the singularity at the center. This texture has a nonvanishing vorticity $n$ when $\gamma = n \phi$. This spin texture has been observed as a coreless vortex composed of three-component order parameters $\Psi_{\pm 1, 0}$ of a spinor BEC \citep{Leanhardt2}.

\subsubsection{Stimulated Raman process}\label{stirapsec}
Some papers proposed generating vortices in a BEC using stimulated Raman processes with configurations of optical fields that have orbital angular momentum (OAM) \citep{Marzlin,Marzlin2,Dum,Nandi,Kapale}. A light beam with a phase singularity, such as a Laguerre-Gaussian (LG) beam, has a well-defined OAM along its propagation axis. The set of LG modes 
\begin{equation}
{\rm LG}^{\it l}_{\it p} (r, \phi) = \sqrt{\frac{2 p}{\pi(|l|+p)}} \frac{1}{w_{0}} \left( \frac{\sqrt{2}r}{w_{0}} \right)^{|l|} L_{r}^{|l|}\left( \frac{2r^{2}}{w_{0}^{2}} \right) e^{-r^{2}/w_{0}^{2} + i l \phi} 
\end{equation}
defines a possible basis set to describe paraxial laser beams, where $w_{0}$ is the beam width, $\it{l}$ the winding number, and $\it{p}$ the number of radial nodes for radius $r>0$. Each photon in the $\rm{LG}^{{\it l}}_{{\it p}}$ mode carries OAM $l\hbar$ along its direction of propagation. 

A group at NIST \citep{Anderson3,Ryu} used a 2-photon stimulated Raman process with a Gaussian laser beam propagating along $+x$ and a $\rm{LG}^{1}_{0}$ beam, carrying $\hbar$ of OAM, propagating along $-x$. Interference of counter-propagating Gaussian beams generates a moving sinusoidal optical dipole potential, which can give a directed linear momentum (LM) to Bose-condensed atoms via Bragg diffraction. The potential generated by interference of the counter propagating $\rm{LG}^{1}_{0}$ and Gaussian beams is not sinusoidal but corkscrew-like, due to the radial intensity profile and the helical phase of the $\rm{LG}^{1}_{0}$ beam. Diffraction off this corkscrew potential produces a vortex state with a center-of-mass motion, where atoms that absorb a photon from one beam and simulatedly emit a photon into the other beam acquire both LM and OAM difference of the beams, which in this case was $2 \hbar k$ ($k$ the photon wavevector) and $\hbar$, respectively. 

They generated vortices of higher charge by transferring to each atom the angular momentum from several $\rm{LG}^{1}_{0}$ photons \citep{Anderson3,Ryu}. This experiment directly demonstrated that the OAM of a photon is transferred coherently to an atom in quantized units of $\hbar$. In some situations it might be desirable to generate rotational states with no net LM. This could be accomplished by using an initial Bragg diffraction pulse to put atoms in a non-zero LM state from which they could subsequently be transferred to a rotational state with zero LM. 

\section{A single vortex in an atomic BEC} \label{Singlevortex}
In this section, we concentrate on the problem of a single vortex state in a trapped  BEC. As described in Sec. \ref{vortexbasictheory}, vortex stability is ensured  by the rotation of the system. Studying the motion of a vortex line is the first step towards understanding superfluid hydrodynamics in such a system. Trapped BECs are mesoscopic systems in the sense that the healing length is not significantly smaller than the sample size. Thus, vortex dynamics have noticeable effects on the collective excitation of the condensate, in contrast to the case for traditional superfluid helium systems.

\subsection{Equilibrium properties}
The solution of the GP equation shows that a vortex has a core with a size of the order of the healing length $\xi$, in which the condensate density is zero. When ${\bf \Omega}=\Omega \hat{\bf z}$, vortices are identified by a density dip in the transverse density distribution (in the $xy$ plane). TOF observations by the ENS group \citep{Madison}, however, show that the density is not completely zero in the dip. This result implies that the vortex line is not necessarily straight, because the condensate density along the $z$-axis (rotation axis) is integrated for the transverse image. Surprisingly, such vortex bending remains stationary in the ground state of a cigar-shaped condensate \citep{Garcia-Ripoll,Aftalion,Modugno2,Aftalion2}.

\begin{figure}
\centerline{\includegraphics[height=0.3\textheight]{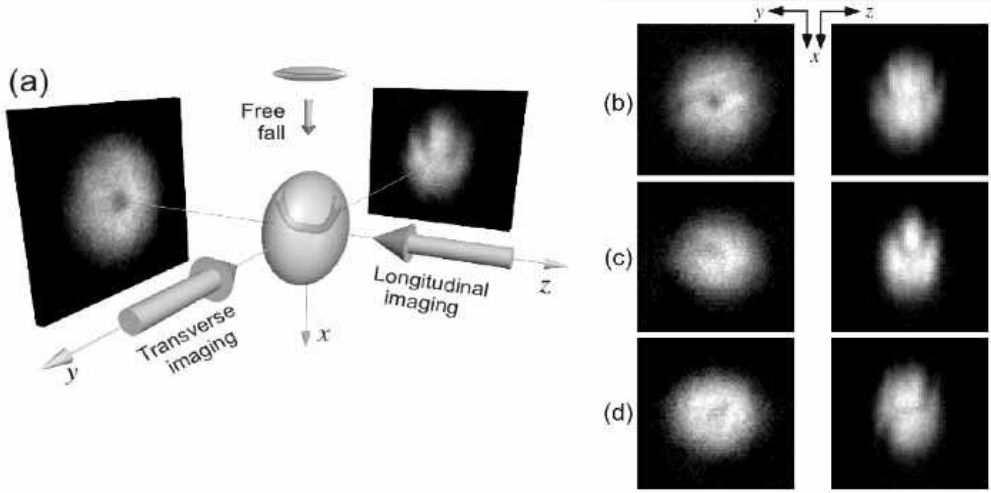}}
\caption{(a) Schematic of the imaging method of a vortex line. Two beams image the atom cloud simultaneously along the longitudinal ($z$) and transverse ($y$) directions of the initial cigar. In (b)--(d), the left column shows the ``longitudinal'' view along $z$, representing the atom distribution in the $xy$ plane. The right column depicts the ``transverse'' view along the $y$ direction, representing the atom distribution in the $xz$ plane. The images were taken after equilibration times of (b) 4 s, (c) 7.5 s, (d) and 5 s. In the atom distribution of the transverse image, the ellipticity is inverted with respect to the initial cigar form, caused by transverse expansion during the TOF. (Taken from \citep{Rosenbusch}. Reprinted with permission from APS.)}
\label{fig:images}
\end{figure}
Evidence of vortex bending in the ground state was observed by the ENS group \citep{Rosenbusch}. They prepared a single vortex state slightly above $\Omega_{c}$ and equilibrated it for a sufficient long time. In the TOF measurements, two imaging beams were aligned along the $y$ and $z$ directions and probed the atom distribution simultaneously [Fig. \ref{fig:images} (a)]. The transverse image in Fig. \ref{fig:images} (b) shows the vortex line, corresponding to the lower atom density; it is not straight and has the shape of a wide ``U''. Figure \ref{fig:images} (c) shows the decay of a U vortex for which the angular momentum has decreased significantly compared to Fig. \ref{fig:images} (b). In the longitudinal view, we can see a vortex off-center. In the transverse view, a narrow U can be seen.

This result is  supported by theoretical analysis based on the 3D GP equation with experimentally appropriate parameters \citep{Garcia-Ripoll,Aftalion,Modugno2,Aftalion2}, where the ground state with a rotation ${\bf \Omega} = \Omega \hat{\bf z}$ was calculated by minimizing the energy functional in Eq. (\ref{GPenergyrot}). The central vortex is generally bent if the trap aspect ratio $\lambda = \omega_{z}/\omega_{\perp}$ is much less than unity. A simple physical picture of the bending can be gained by viewing a cigar-shaped condensate as a series of 2D sheets at various $z$ \citep{Modugno2}. For each sheet, there is a corresponding 2D vortex stability problem with a critical frequency $\Omega^{\rm 2D}_{c}(z)$ (see Sec. \ref{vortexbasictheory}) above which a centered vortex is the stable solution. Since the effective 2D chemical potential is $\mu_{\rm 2D}(z) = \mu - m \omega_{z}^{2} z^{2} / 2$, the radius of the 2D condensate at $z$ becomes $R_{\perp}^{2}(z) = 2 \mu_{2D}(z)/m\omega_{\perp}^{2} = R_{\perp}^{2}(0) - \lambda^{2} z^{2}$. Thus, $\Omega^{\rm 2D}_{c}(z) \propto R_{\perp}$ is a decreasing function from $z=0$ to $|z|=R_{z}$. For a given rotation frequency $\Omega$, the vortex line minimizing the total energy is well centered for $|z|<z_{c}$ and is strongly bent for $|z|>z_{c}$, where $\Omega^{\rm 2D}_{c}(z_{c})=\Omega$. This bending is a symmetry-breaking effect, which does not depend on the presence of rotating anisotropy and which occurs even in a completely axisymmetric system \citep{Garcia-Ripoll}. A precursor of this bending effect can be found in the excitation spectrum of a condensate with a centered straight vortex \citep{Svidzinsky,Feder}, in which negative-energy modes localized at the core (so-called ``anomalous modes'') appear with increasing $\lambda$. As these modes grow, the central vortex is pushed outward. This indicates that the bending instability needs a dissipation mechanism and that it takes a long time at low temperatures.

Figure \ref{fig:images} (d) shows a vortex line in the shape of an ``S'', which can be regarded as a U vortex with a half part rotated by 180$^\circ$. An S single vortex can also be found by numerical simulation as the stationary state of an elongated condensate for a given rotation frequency \citep{Aftalion2}, having an energy higher than that of the U vortex \citep{Komineas}.

\subsection{Dynamical properties}
\subsubsection{Precession and decay of an off-centered vortex}
Precession of a vortex core off-center in a condensate is a simple example of vortex motion. Core precession can be described in terms of a Magnus force effect. A net force on a quantized vortex core creates a pressure imbalance, resulting in core motion perpendicular to both the force and the vortex quantization axis. In the case of trapped BECs, these net forces can be caused by either condensate density gradients \citep{Svidzinsky2,Svidzinsky,Jackson,McGee} or drag due to thermal atoms \citep{Fedichev}. The former may be thought of as a sort of effective buoyancy. Typically, the total buoyancy force is towards the condensate surface and the net effect is a precession of the core around the condensate axis via the Magnus effect. The latter causes radial drag and spiraling of the core towards the condensate surface due to energy dissipation and damping processes.

Core precession has been investigated in detail by the JILA group \citep{Anderson1}. Starting with a composite vortex created by the phase-engineering method of Sec. \ref{phaseeng}, they selectively removed components filling the vortex core with resonant light pressure. In the limit of complete removal, a single-component vortex state with a bare core can be obtained. The vortex was off-center because of the instability of the formation process. The precession frequency was determined from the vortex position taken directly from the density profile. The vortex core precessed in the same direction as the vortex fluid flow around the core. The obtained result of 1.8 Hz agrees well with analytical results based on the Magnus force picture \citep{Svidzinsky2,Svidzinsky} and more precise numerical simulations \citep{Jackson,McGee,Feder}.

In some results, the vortex core disappeared from the observed images during the time evolution. However, these were not associated with the decay of vortices because there was no evidence of radial spiral motion due to energy dissipation, which could be caused by the thermal drag \citep{Fedichev} or the sound radiation from a moving vortex \citep{Lundh0,Parker1}. Subsequently, an experiment using the surface-wave spectroscopic technique revealed that vortices were actually present for a long time in the condensate \citep{Haljan}. The main cause of the disappearance was the tilting motion of a vortex, the lowest odd-order normal mode of a single-vortex state, which is sensitive to small trap anisotropy \citep{Svidzinsky}.

\subsubsection{Vortex dynamics coupled with collective modes}\label{gyroscope}
Vortex dynamics are greatly affected by the overall collective motion of the condensate because of the mesoscopic nature of the system. Here, we detail several interesting results of coupled dynamics. The determination of the frequency of the collective modes allows precise measurement of the angular momentum carried by the quantized vortices.
\\

(i) {\it Transverse quadrupole mode}

The collective modes of a trapped condensate can be classified by expressing the density fluctuations in terms of polynomials of degree in the Cartesian coordinates $(x_{1},x_{2},x_{3}) = (x,y,z)$. The quadrupole modes are characterized by a density fluctuation with a polynomial of second order, e.g., $\delta n = \sum p_{ij} x_{i} x_{j}$, which gives six normal modes. In an axisymmetric harmonic potential, linear combinations of the diagonal components $p_{xx}$, $p_{yy}$, and $p_{zz}$ describe three normal modes: one transverse mode with $m_{z}=2$ and two radial-breathing modes with $m_{z}=0$, where $m_{z}$ is the projected angular momentum on the symmetry axis. The remaining three normal modes are associated with the off-diagonal components $p_{xy}$, $p_{yz}$, and $p_{zx}$, which are scissors modes \citep{Odelin,Marago}.

The first study was done for the excitation of two transverse quadrupole modes with $m_{z}= \pm 2$ \citep{Chevy}. When a vortex is present in a condensate, the frequencies of the $m_{z}= \pm 2$ quadrupole modes increase by $\omega_{+} - \omega_{-} =2 \langle L_{z} \rangle /m N \langle x^{2}+y^{2} \rangle$ because of broken rotational symmetry \citep{Zambelli}, where $\langle \hspace{2mm} \rangle$ stands for the average within the condensate. The increase causes precession of the eigenaxes of the quadrupole mode at an angular frequency $\dot{\theta} = (\omega_{+}-\omega_{-})/2|m_{z}|$. By measuring the angular velocity of this precession, we can determine the mean angular momentum $\langle L_{z} \rangle$ of the condensate. This spectroscopic method has also been used to characterize the tilting motion of a vortex \citep{Haljan} and the winding number of a single vortex \citep{Leanhardt}.

Excitation of the transverse quadrupole mode yields further interesting vortex dynamics. The ENS group observed that when the superposition of the $m_{z}=\pm2$ quadrupole modes are excited with equal amplitudes, the oscillation of the $m_{z}=-2$ mode decays faster than that of the $m_{z}=+2$ mode \citep{Bretin}. A possible physical origin of this phenomenon is that the $m_{z}=-2$ mode decays to Kelvin modes through a non-linear Beliaev process. This is supported by theoretical analysis based on the BdG equation \citep{Mizushima}. Kelvin modes correspond to long-wavelength helical traveling waves along a vortex line with a dispersion relation $\omega_{\rm K} \simeq (\hbar k^{2}/2m) \ln(1/k\xi)$ $(k\xi \ll 1)$. According to the Kelvin--Helmoltz theorem, the Kelvin modes rotate always in the sense opposite to the vortex velocity field. Consequently, the angular momentum of a quantum of a Kelvin mode associated with a singly-quantized vortex is $-\hbar$. Because of the negative angular momentum with respect to the vortex winding number, this mechanism is effective only for the $m_{z}=-2$ mode.  From the energy $\omega_{-2}=2\omega_{\rm K}$ and angular momentum conservation, an excitation of the quadrupole mode $m_{z}=-2$ can decay to a pair of Kelvin waves with wave vectors $k$ and $-k$, while angular momentum conservation forbids the decay of the $m_{z}=+2$ mode.
\\

(ii) {\it Gyroscope motion}

\begin{figure}[hbtp]
\begin{center}
\includegraphics[height=0.16\textheight]{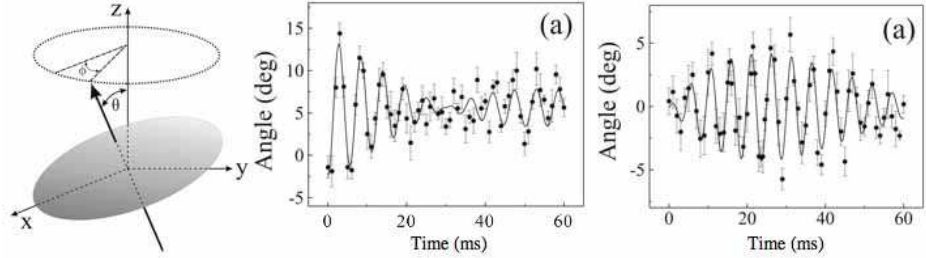}
\end{center}
\caption{Left: schematic picture of gyroscope motion of the experiment of Ref. \citep{Hodby2}. The scissors mode involves a fast oscillation of the small angle $\theta$ between the condensate normal axis and the $z$ axis. When a vortex is present, the plane of this oscillation (initially the $xz$ plane with $\phi=0$) slowly precesses through angle $\phi$ about the $z$ axis. Right: data of the evolution of the tilt angle $\theta$ projected onto the $xz$ plane, when the scissors mode is initially excited (a) in the $xz$ plane and (b) in the $yz$ plane. (Taken from \citep{Hodby2}. Reprinted with permission from APS.)}
\label{Oxfordgyro}
\end{figure}

What happens when the other quadrupole modes are excited in a condensate with a vortex? The Oxford group studied the response of a condensate with a vortex when the $xz$ or $yz$ scissors modes are excited \citep{Hodby2}. Similar to the case of transverse quadruople modes, in the presence of the vortex, the plane of oscillation of a scissors mode precesses slowly around the $z$ axis. In polar coordinates, the scissors oscillation is in the $\theta$ direction and the precession is in the $\phi$ direction, as shown schematically in Fig. \ref{Oxfordgyro}. This can be regarded as a kind of gyroscope motion of the vortex line \citep{Stringari}.

The relationship between the precession rate and $\langle L_z \rangle$ can be derived by considering the scissors mode as an equal superposition of two counter-rotating $m_{z} = \pm 1$ modes. These modes represent a condensate tilted by a small angle from the horizontal plane rotating around the $z$ axis at the frequency of the scissors oscillation, $\omega_{\pm}$ = $\omega_{sc}$. The symmetry and degeneracy of these modes are also broken by the axial angular momentum $\langle L_z \rangle $. By applying a similar argument as that for transverse quadrupole modes \citep{Stringari}, the precession rate is related to the frequency splitting,  $\omega_{+} - \omega_{-} = \langle L_z \rangle / m N \langle x^2 + z^2 \rangle$, allowing the angular momentum $\langle L_{z} \rangle$ to be determined.

The precession associated with gyroscope motion can be seen in the results of Fig. \ref{Oxfordgyro}(a) and (b), corresponding to a slowly varying oscillation component. The pattern of increase and decrease of the amplitude is exchanged between (a) and (b), with different directions of the excitation. This is clear evidence of a slow precession along the $\phi$-direction. The results also show that the motion of the vortex core exactly follows the axis of the condensate. These observations can be well reproduced by direct numerical simulations of the 3D GP equation \citep{Nilsen}. From the precession rate, the measured angular momentum per particle associated with a vortex line was found to be $1.07 \hbar \pm 0.18 \hbar$.

\subsubsection{Splitting of a multiply quantized vortex}\label{splittingbec}
The energy cost to create a $q>1$ vortex is  less favorable than $q$ single-quantized vortices, as seen in Eq. (\ref{onevorenrgy}). This raises an interesting question as to what happens when such an unstable vortex is created. As it happens, topological phase imprinting, shown in Sec. \ref{topologicalphase}, can be used to create an unstable vortex.

The stability characteristics of a multiply quantized vortex in a trapped BEC exhibit an interesting interaction dependence because of the finite size effect \citep{Pu}. BdG analysis for a cylindrical system reveals complex eigenvalue modes, which implies that a multiply quantized vortex is dynamically unstable. For a vortex with a winding number $q$, angular momentum conservation leads to constraints on the normal mode functions $u_{j} ({\bf r}) = u_{j} (r) e^{i (\kappa_{j} + q) \theta}$ and $v_{j} ({\bf r}) = v_{j} (r) e^{i (\kappa_{j} - q) \theta}$, where $\kappa_{j}$ denotes the angular momentum quantum number of the mode. For $q=2$ there are alternating stable and unstable regions with respect to the interaction parameter $a n_{z} = a \int |\Psi|^{2} dx dy$; the first and second regions appears for $0 < an_{z} < 3$ and $11.4 < an_{z} < 16$. Numerical simulations demonstrate that when a system is in an unstable region, a doubly quantized vortex decays into two singly quantized vortices \citep{Mottonen}.

An experiment by the MIT group studied the splitting process of a doubly quantized vortex and its characteristic time scale as a function of $an_{z=0}$ \citep{Shin}. The results show that a doubly quantized vortex decays, but that the lifetime increases monotonically with $an_{z=0}$, showing no periodic behavior. This contrary to the above theoretical prediction. Recent numerical studies of 3D GP equations reveal this mysterious observation, emphasizing that the detailed dynamical behavior of a vortex along the entire $z$-axis is relevant for characterizing the splitting process in an elongated condensate \citep{Huhtamaki2,Mateo}.

\begin{figure}
\begin{center}
\includegraphics[height=0.26\textheight]{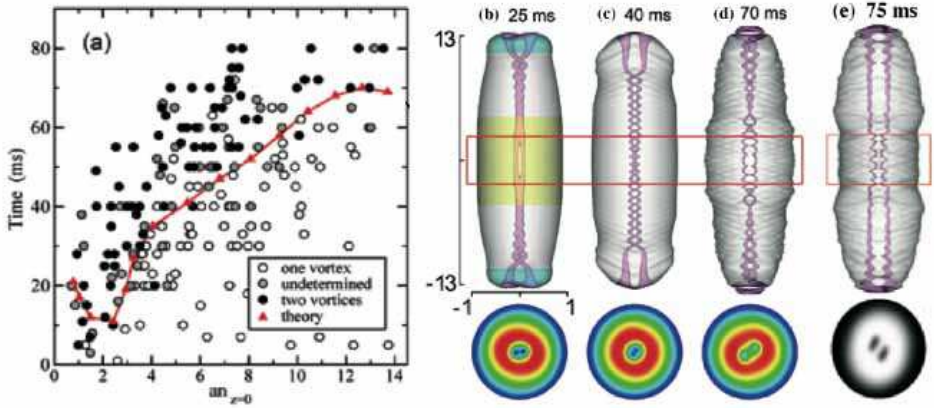}
\end{center}
\caption{Splitting process of a doubly quantized vortex. (a) Experimental data combined with theoretically predicted splitting times as a function of $an_{z=0}$ for a 4\% quadrupolar perturbation acting during 0.3 ms. The splitting is identified by the number of visible vortex cores from the density profile of the axial absorption images taken from a 30-$\mu$m thick central slice of the condensate. (b)--(e) Time evolution of the splitting process for $an_{z=0} = 13.75$, obtained by a 3D simulation of the GP equation. The shaded zones in (b) indicate the instability regions. The corresponding axial absorption images of the central slice are also shown at the bottom of the figure. The lengths are in units of 6.05 $\mu$m. (Taken from \citep{Mateo}. Reprinted with permission from APS.)}
\label{splittingvor}
\end{figure}

The trigger for splitting instability is likely to be gravitational sag during the formation process with a reversing axial bias field $B_{z}$ \citep{Huhtamaki2}. In experiments, absorption images were restricted to a 30-$\mu$m thick central slice of the condensate to increase the visibility of the vortex cores. The experimental results in Fig. \ref{splittingvor}(a) shows that the fastest decay occurs at $an_{z} \simeq 1.5$, consistent with the BdG and numerical analysis. As the particle number increases, the first instability region ($0<an_{z}<3$) moves progressively away from the central slice toward the edges of the condensate because of the trapping potential. As a consequence, the splitting instability of the vortex core has to propagate from those regions to the center. This process is responsible for the monotonic increase in the lifetime for $an_{z} > 3$. According to the theory, a second minimum is expected about $an_{z=0} \simeq 13.75$. Even though no such minimum occurs, a change in the slope of the predicted curve at $an_{z=0} \simeq 13.75$ is seen as $an_{z=0}$ enters the second instability region \citep{Huhtamaki2,Mateo}. Figures \ref{splittingvor} (b)--(e) show the time evolution of the splitting process for $an_{z=0}=13.75$. The first and second instability regions correspond to the shaded zones in Fig. \ref{splittingvor} (b). This clearly shows that at $t=25$ ms, the splitting process has already begun in both the edges and the center of the condensate. The different precession frequency along different $z$ slices causes inter-winding of two single-quantized vortices. However, the two vortex cores near the central slice still overlap (Fig. \ref{splittingvor} (c)) and thus cannot be experimentally resolved until much longer times. At $t \approx 70$ ms (Fig. \ref{splittingvor} (d)), the vortex cores begin to disentangle so that they can be unambiguously resolved at $t=75$ ms (Fig. \ref{splittingvor}(e)). Thus, there is no contradiction with the theoretical prediction.

The physical origin of the periodic appearance of the unstable region is anomalous modes with negative eigenvalues. When the eigenvalue of positive-energy modes coincides with the absolute value of the eigenvalue of negative-energy modes, a complex eigenvalue mode are produced through mutual annihilation of these two excitations \citep{Skryabin, Kawaguchi,Jackson11,Lundh4}; the total angular momentum of these excitations is also vanished. Since the $an_{z}$ dependence of the negative-energy eigenvalues is very different from that of positive-energy ones, the above matching condition can be satisfied in sequence with increasing $an_{z}$, which results in the periodic appearance of the complex-eigenvalue modes. Thus, splitting instability of a multiply quantized vortex can be suppressed for a particular trap asymmetry and interaction strength because the collective excitations depend strongly on the character of the trapping potential \citep{Huhtamaki}. It has also been predicted that multiply quantized vortices can be stabilized by introducing a suitable localized pinning potential \citep{Simula} or non-simply connected geometry such as quartic confinement \citep{Lundh2}. Very recently, splitting dynamics of a quadruply quantized ($q=4$) vortex was reported \citep{Isoshima3}.

\section{A lattice of quantized vortices in an atomic BEC} \label{Vortexlattice}
We now address the issue of a rapidly rotating BEC where many vortices have been nucleated and arranged into a regular triangular lattice \citep{Abo,Coddington2}. We first present a discussion of the equilibrium properties of a rapidly rotating condensate, and then present the basic theoretical background for its description. The equilibrium properties of vortex lattices in a trapped BEC have been extensively studied by the JILA group \citep{Schweikhard,Coddington2}. We next discuss the collective dynamics of an assembly of vortices in a trapped BEC. Finally, we discuss an unconventional vortex phase which occurs in the presence of an externally applied potential created by laser beams.

\subsection{Equilibrium properties}\label{equilibriumprop}
For very large $\Omega$, the rotation of the superfluid mimics a rigid body rotation with $\nabla \times {\bf v}_{\rm s} = 2 {\bf \Omega}$ by forming a vortex lattice. Using the fact that the vorticity is given by the form $\nabla \times {\bf v}_{s} = \kappa \delta^{(2)} ({\bf r}_{\perp}) \hat{\bf z}$, we find that the average vorticity per unit area is given by $\nabla \times {\bf v}_{\rm s} = \kappa n_{v} \hat{\bf z}$, where $n_{v}$ is the number of vortices per unit area. Hence, the density of the vortices is related to the rotation frequency $\Omega$ as $n_{v} = 2 \Omega/\kappa$ \citep{Feynman}. This relation can be used to estimate the maximum possible number of vortices in a given area as a function of $\Omega$. As shown below, the properties of a vortex lattice can be characterized by the nearest-neighbor lattice spacing $\sim b=(\hbar/m\Omega)^{1/2}$, defined by the area per vortex $n_{v}^{-1} = \pi b^{2}$, and by the radius of each vortex core $\sim \xi$.

Note that the GP energy functional of Eq. (\ref{GPenergyrot}) in a rotating frame can be rewritten as
\begin{equation}
E' = \int d {\bf r} \left( \frac{\hbar^{2}}{2m} \left| \left(- i \nabla - \frac{m}{\hbar} {\bf \Omega} \times  {\bf r} \right) \Psi \right|^{2} + V_{\rm eff} |\Psi|^{2} + \frac{g}{2} |\Psi|^{4} \right), \label{GPenergyrot2}
\end{equation}
where $V_{\rm eff} = m (\omega_{\perp}^{2} - \Omega^{2}) r^{2} /2 + m \omega_{z}^{2} z^{2} / 2$ is the effective trapping potential combined with the centrifugal potential; the rotation effectively softens the radial potential and vanishes at $\Omega = \omega_{\perp}$. Because the first term in Eq. (\ref{GPenergyrot2}) reads $\hbar^{2} (\nabla |\Psi|)^{2} / 2m + m ({\bf v}_{s} - {\bf \Omega} \times {\bf r})^{2} |\Psi|^{2} /2 $, it can be neglected in the TF limit and the rigid-body rotation limit ${\bf v}_{s} = {\bf \Omega} \times {\bf r}$. Then, the TF radius is given by $R_{\perp}(\Omega) = R_{\perp} / [1-(\Omega/\omega_{\perp})^{2}]^{3/10}$ with $R_{\perp}$ for nonrotating condensate and an aspect ratio of $\lambda_{\rm rb} = R_{\perp}(\Omega)/R_{z} = \lambda / [1-(\Omega/\omega_{\perp})^{2}]^{1/2}$. Thus, measuring $\lambda_{\rm rb}$ will give the rotation rate of the condensate \citep{Raman2,Haljan}. Also, in the high rotation limit $\Omega \rightarrow \omega_{\perp}$, the condensate flattens out and reaches an interesting quasi-2D regime; current experiments have reached $\Omega/\omega_\perp\approx 0.995$ \citep{Coddington2}.

Abrikosov triangular lattice of quantized vortices have been observed experimentally, as shown in Fig. \ref{MITvordenang}. Direct imaging allows a detailed investigation of the nature of a vortex lattice in an inhomogeneous superfluid. Here, we summarize the salient results and theoretical interpretations of these observations.

\subsubsection{Lattice inhomogeneity}\label{latticeinhomo}
Experimental observations \citep{Abo,Engels} and numerical simulation of the 3D GP equation \citep{Feder4} have revealed that for a finite-size trapped BEC, the vortex density in a lattice is lower than the rigid-body estimate and the lattice is remarkably regular. Sheehy and Radzihovsky explained these points analytically in the TF limit \citep{Sheehy1,Sheehy2}; they derived a small, radial-position-dependent, inhomogeneity-induced correction term to the vortex density as
\begin{equation}
n_{v} (r)= \frac{2 \Omega}{\kappa} - \frac{1}{2 \pi} \frac{R_{\perp}(\Omega)^{2}}{ (R_{\perp}(\Omega)^{2} - r^{2})^{2} } \ln \frac{e^{-1} \kappa}{2 \pi \Omega \xi^{2}}. \label{rigidcorrection}
\end{equation}
This result indicates that the vortex density is always lower than the rigid-body estimate of the first term in Eq. (\ref{rigidcorrection}). Also, the vortex density is higher in regions where the condensate density is most uniform, that is, the central part of a harmonically trapped gas. However, the position-dependent correction is small ($n_{v}$ changes less than a few \% over a region in which the atom density varies by 35\%), which seemingly causes regularity of the lattice. These results have been confirmed by a detailed experimental study \citep{Coddington2}. It should be also noted that inhomogeneity in the area density of vortices can also be derived in the limit of the lowest Landau level \citep{Watanabe,Cooper1,Aftalion4,Baym4}, as explained below.
\\

\subsubsection{Attainment of the lowest Landau level regime}
Note that the first term in Eq. (\ref{GPenergyrot2}) can be identified as the Hamiltonian $H_{L} = (-i \hbar \nabla -e {\bf A}/c)^{2}/2m$ of a charge $-e$ particle moving in the $xy$ plane under a magnetic field $B \hat{\bf z}$ with a vector potential ${\bf A} = (mc/e) {\bf \Omega} \times {\bf r}$. If the interaction is neglected ($g=0$), the eigenvalues of the Hamiltonian of Eq. (\ref{GPenergyrot2}) forms Landau levels as $\epsilon_{n, m} / \hbar = \omega_{\perp} +  n (\omega_{\perp} + \Omega) +m (\omega_{\perp} - \Omega)$, where $n$ is the Landau level index and $m$ labels the degenerate states within a Landau level. The lowest energy states of two adjacent Landau levels are separated by $\hbar(\omega_\perp + \Omega)$, whereas the distance between two adjacent states in a given Landau level is $\hbar(\omega_{\perp} - \Omega)$; when $\Omega = \omega_\perp$, all states in a given Landau level are degenerate. Physically, this corresponds to the case where the centrifugal force exactly balances the trapping force in the $x$-$y$ plane, and only the Coriolis force remains. The system is then invariant under translation and hence has macroscopic degeneracy. This formal analogy has led to the prediction that quantum Hall-like properties would emerge in rapidly rotating BECs \citep{Ho,Viefers,Cooper2,Sinova,Regnault,Regnault2,Cazalilla,Chang2,Rezayi,Morris}.

Interaction effects mix different ($m$, $n$) states. Because the density $\bar{n}$ of the system drops as $\Omega\to \omega_\perp$, the interaction energy $\sim g \bar{n}$ can become small compared with the Landau level separation $2 \hbar \omega_{\perp}$. In this limit, particles should condense into the lowest Landau level (LLL) with $n=0$. Then, the system enters the ``mean field'' quantum Hall regime, where the wave function can be described by only the LLL orbitals with the form $\Psi_{\rm LLL} = A \prod_{j}( z - z_{j}) e^{-r^2/2 a_{\rm ho}^2}$, where $z=x+iy$, $z_{j}$ is the positions of vortices (zeros), and $A$ is a normalization constant. The minimization of Eq. (\ref{GPenergyrot2}) using the ansatz $\Psi_{\rm LLL}$ is a useful theoretical prescription to tackle the properties of rapidly rotating BECs \citep{Watanabe,Cooper1,Aftalion4,Sonin2,Aftalion5,Cozzini2}.

Schwaikhard {\it et al}. created rapidly rotating BECs by spinning condensates to $\Omega/\omega_{\perp} > 0.99$ \citep{Schweikhard}. When the condensate enters the LLL regime, characteristic equilibrium properties appear, as described below.
\\

(i) {\it Global structure}

In the LLL limit, the radial condensate density profile with uniform vortex distribution has been predicted to change from a parabolic TF profile to a Gaussian profile as $\Omega$ is increased \citep{Ho}. However, no signs of such crossover have been found in experiments; even when the dynamics were restricted to the LLL, the density profile remained a parabolic TF profile as $\Omega \to \omega_\perp$ \citep{Schweikhard,Coddington2}. This result can be seen qualitatively from the energy minimization under the LLL limit with nonuniform vortex density \citep{Watanabe,Cooper1,Aftalion4}. As long as the total number of vortices is much larger than unity ($N_{v} \gg 1$), the energy in the LLL regime is given by
\begin{equation}
E'= \Omega N +\int d {\bf r} \left[ (\omega_{\perp}-\Omega) \frac{r^2}{a_{\rm ho}^2}n({\bf r}) +\frac{bg}{2} n({\bf r})^2 \right],
\label{LLLenergy}
\end{equation}
plus terms involving the trapping potential in the $z$-direction. Here, $n({\bf r}) = \langle |\Psi_{\rm LLL}|^2\rangle$ is the coarse-grain averaged density profile in order to smooth the rapid variations at the vortex cores. Then, the interaction parameter $g$ is renormalized to $bg$, where $b = \langle |\Psi_{\rm LLL}|^4\rangle / \langle |\Psi_{\rm LLL}|^2\rangle^{2}$ is the Abrikosov parameter. The energy (\ref{LLLenergy}) is then minimized by the TF profile, $n(r) = [\mu - \Omega -(\omega_{\perp} - \Omega)  r^2/a_{\rm ho}^2] / bg $.

Since the energy (\ref{LLLenergy}) depends only on the smoothed density, the vortices adjust their locations so that the smoothed density becomes an inverted parabola. In the LLL regime, the relation between the condensate density $n({\bf r})$ and the mean vortex density, $n_{v} ({\bf r}) = \sum_{j} \delta({\bf r} - {\bf r}_{j})$ is given by $4^{-1}\nabla^2 \ln n ({\bf r})  = - a_{\rm ho}^{-2} +\pi n_v ({\bf r})$ \citep{Ho,Aftalion4}. If the density profile is Gaussian, the vortex density is constant.  However, for a TF profile,
\begin{equation}
n_v (r) = \frac{2 \Omega}{\kappa} - \frac{1}{\pi R_{\perp}^2}\frac{1}{\left(1 - r^2 / R_{\perp}^2 \right)^2}. \label{nnv}
\end{equation}
This result is similar to Eq. (\ref{rigidcorrection}) in the low rotation regime, where the coefficient of the second term is different. Since the second term is smaller than the first, as $\sim a_{\rm ho}^{2}/R_{\perp}^{2} \simeq N_{v}^{-1}$, the density of the vortex lattice is basically uniform, consistent with the argument in Sec. \ref{latticeinhomo}. Turning the argument around, very small distortions of the vortex lattice from perfect triangular can result in large changes in the global density distribution such that the TF form is energetically favored rather than the Gaussian. 
\\

(ii) {\it Vortex core structure and fractional area}

\begin{figure}[hbtp]
\begin{center}
\includegraphics[height=0.25\textheight]{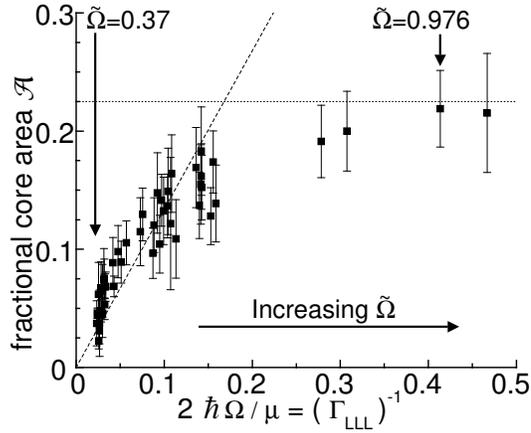}
\end{center}
\caption{Fraction of the condensate surface area occupied by vortex cores ${\cal A}$ (see text) versus the inverse of the LLL parameter $\Gamma_{\rm LLL}^{-1}$, measured after condensate expansion. The data clearly show a saturation of ${\cal A}$ as $\Omega/\omega_{\perp} \rightarrow 1$. The dashed line represents the prediction for the pre-expansion value in the case of a low rotation rate. The dotted line shows the results for a saturated value of ${\cal A}$ in the LLL limit. (Taken from \citep{Schweikhard}. Reprinted with permission from APS.)}
\label{corearea}
\end{figure}
Another interesting characteristic of the LLL is that the vortex core is of the same size as the distance $b=(\hbar/m\Omega)^{1/2}$ between adjacent vortices. The radius of a single vortex core is of order $\xi = 1/\sqrt{8\pi n a}$, so that a vortex core would begin to overlap the next at $\xi \sim b$. This gives an upper-critical rotation frequency $\Omega_{c1} \sim 8\pi n a \hbar/m \sim 10^3 - 10^5$ rad/sec, which is an experimentally accessible rate. However, there is no phase transition associated with vortex cores overlapping in a rotating condensate. Rather, vortex cores begin to shrink as the intervortex spacing becomes comparable to the healing length $\xi$, and eventually the core radius scales with the intervortex spacing \citep{Fischer,Baym3,Watanabe2}. 

Figure \ref{corearea} shows the measured fractional area, defined as ${\cal A} = r_{v}^{2}/b^{2} = n_{v} \pi r_{v}^{2}$, as a function of the inverse of the LLL parameter $\Gamma_{\rm LLL} = \mu / 2 \hbar \Omega$ \citep{Schweikhard,Coddington2}. The linear rise of ${\cal A}$ at small $\Omega$ occurs because the core size remains constant, while $n_{v}$ increases linearly with $\Omega$. Explicitly, the core radius was estimated numerically as $r_{v} = 1.94 \xi$ \citep{Schweikhard}, and $n_{v} = m \Omega/\pi  \hbar$ by neglecting the effect of inhomogeneity. These values yield ${\cal A} = 1.34 \Gamma_{\rm LLL}^{-1}$, shown by the dashed line in Fig. \ref{corearea}, where $n = 0.7 n_{\rm peak}$ and $\mu = g n_{\rm peak}$ were used for the estimation. The flattening of ${\cal A}$ with increasing $\Omega$ is a consequence of the vortex radius scaling with the intervortex spacing. The upper dotted line shows the upper limit ${\cal A}=0.225$, obtained by using the oscillator $p$-state structure $|\Psi_{\rm core}(r)|^{2} \sim [(r/b) \exp (-r^{2}/2b^{2})]^{2}$ as the profile of a vortex core in the LLL limit, where $b$ is regarded as the radius of the (cylindrical) Wigner--Seitz cell around a given vortex. The data in Fig. \ref{corearea} show the expected initial linear rise, with the predicted scaling of the core radius with intervortex spacing. More detailed theoretical studies which treat the core structure explicitly obtain excellent agreement with the experimental results \citep{Cozzini2,Watanabe2}.

\subsection{Collective dynamics of a vortex lattice} \label{collectivedynvorlat}
\subsubsection{Vortex lattice dynamics coupled with collective modes}
As shown in Sec. \ref{gyroscope}, vortex states undergo interesting responses to excitation of the transverse quadrupole modes with $m_{z} = \pm 2$. Similar studies have been made for rapidly rotating BECs. In this case, the dispersion relation of the quadrupole modes is given by $\omega_{\pm 2} = \sqrt{2 \omega_{\perp}^{2} - \Omega^{2}} \pm \Omega$ \citep{Cozzini0}, which has been measured experimentally \citep{Haljan}. When $\Omega \rightarrow \omega_{\perp}$, we have $\omega_{+2} \rightarrow 2 \omega_{\perp}$ and $\omega_{-2} \rightarrow 0$, reflecting the tendency of the system to become unstable against quadrupole deformation. Excitation of the quadrupole mode for $\Omega \rightarrow \omega_{\perp}$ induces large deformations of the condensate and nonequilibruim dynamics of vortex lattices \citep{Engels}. Interestingly, when the $m_{z} = -2$ mode was excited, a vortex lattice was distorted to form a one-dimensional set of closely spaced vortices. This observation was explained by the fact that vortices should follow the stream line of the background quadrupole velocity field \citep{Cozzini0,Mueller}. In contrast, excitation of an $m_{z} = +2$ mode dissolved the regular lattice, where the vortex lines were randomly arranged in the $x$-$y$ plane but were still strictly parallel along the $z$-axis. 

As stated in Sec. \ref{equilibriumprop}, a centrifugal force distorts the cloud into an extremely oblate shape, and thus the rotating cloud approaches the quasi-2D regime. Excitation of an axial breathing mode ($m_{z}=0$) has been used to confirm the 2D signature of a rapidly rotating BEC \citep{Schweikhard}. For a BEC in the axial TF regime, an axial breathing frequency $\omega_{\rm B} = \sqrt{3} \omega_{z}$ has been predicted in the limit $\Omega/\omega_{\perp} \rightarrow 1$ \citep{Cozzini0}, whereas $\omega_{\rm B} =2 \omega_{z}$ is expected for a noninteracting gas, expected for $\mu < \hbar \omega_{z}$. Schweikhard {\it et al}. observed a crossover of $\omega_{\rm B}$ from $\sqrt{3} \omega_{z}$ to $2 \omega_{z}$ with increasing $\Omega$ ($\mu \sim 3 \hbar \omega_{z}$). Also, excitation of the scissors mode in a condensate with a vortex lattice induces a collective tilting mode of the vortex array (the lowest-energy Kelvin wave of the lattice) \citep{Smith}, referred to as an anomalous scissors mode \citep{Chevy3}.

\subsubsection{Transverse oscillation of a vortex lattice: Tkachenko mode}
The dynamics of vortex lattices itself raises many interesting problems. It should be possible to propagate collective waves in a transverse direction to the vortex lattice in the superfluid, called Tkachenko (TK) modes. For an incompressible superfluid, the dispersion law is given by $\omega_{\rm TK} (k) = \sqrt{\hbar \Omega/4m} k$. The TK modes of a vortex lattice in a trapped BEC have been analyzed theoretically \citep{Baym2,Mizushima2,Baksmaty,Baym1,Cozzini,Gifford,Sonin,Sonin2} and observed experimentally \citep{Coddington}.

\begin{figure}[hbtp]
\begin{center}
\includegraphics[height=0.20\textheight]{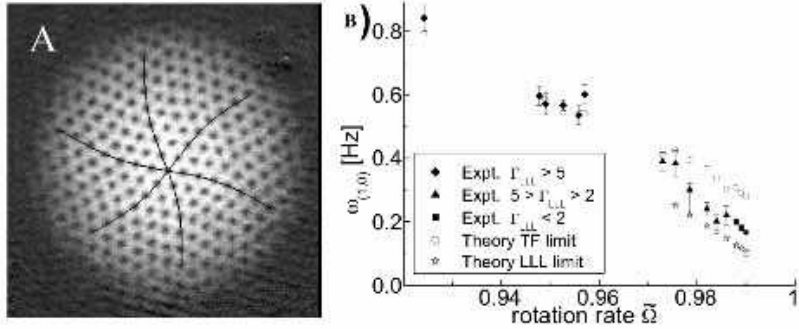}
\end{center}
\caption{(A) TK mode $(n,m)=(1,0)$ at $N=1.5 \times 10^{5}$ and $\Omega = 0.989 \omega_{\perp}$, fitted by sine fits. (B) Comparison of measured TK mode frequency $\omega_{(1,0)}$ (solid symbols) versus the theoretical value \citep{Baym2}, using the vortex lattice shear modulus $C_{2}^{\rm TF}$ in the TF limit (circles) and $C_{2}^{\rm LLL}$ in the LLL regime (stars). Note that both $N$ and $\Gamma_{\rm LLL}$ decrease as $\tilde{\Omega} = \Omega/\omega_{\perp}$ increases. For $\Gamma_{\rm LLL} \simeq 3$ (reached at $N=7.8 \times 10^{5}$ and $\tilde{\Omega} \simeq 0.978$) the data cross over from the TF to the LLL prediction. (Taken from \citep{Schweikhard}. Reprinted with permission from APS.)}
\label{TKmode}
\end{figure}

Experimentally, TK modes have been excited by the selective removal of atoms at the center of a condensate with a resonant focused laser beam, or by the insertion of a red-detuned optical potential at the center to draw atoms into the middle of the condensate. The former method has also been used to create long-lived vortex aggregates \citep{Engels2}. In the experiment, the TK modes were identified by the sinusoidal displacement of the vortex cores with the origin at the center of the condensate; see Fig. \ref{TKmode}(A). TK modes can be classified by the quantum number $(n,m)$, associated with radial and angular nodes, in a presumed quasi-2D geometry.

To explain the observed frequency of the TK mode $\omega_{(n,m)}$, the effects of compressibility should be taken into account. According to the elastohydrodynamic approach developed by Baym \citep{Baym2}, the TK frequency is described by the compressional modulus $C_{1}$ and shear modulus $C_{2}$ of the vortex lattice, included in the elastic energy 
\begin{equation}
E_{\rm el} = \int d {\bf r} \left\{ 2 C_{1} (\nabla \cdot {\bf \epsilon})^{2} + C_{2} \left[ \left( \frac{\partial \epsilon_{x}}{\partial x} - \frac{\partial \epsilon_{y}}{\partial y} \right)^{2} + \left( \frac{\partial \epsilon_{x}}{\partial y} + \frac{\partial \epsilon_{y}}{\partial x} \right)^{2} \right] \right\},
\end{equation}
where ${\bf \epsilon}({\bf r},t)$ is the continuum displacement field of the vortices from their home positions. In the incompressible TF regime, $C_{2}=-C_{1}=n \hbar \Omega/8$. Then, the upper branch of the energy spectrum follows the dispersion law $\omega_{+}^2 = 4 \Omega^{2} + c^{2} k^{2}$ with sound velocity $c=\sqrt{gn/m}$, being the standard inertial mode of a rotating fluid and having a gap at $k=0$. Conversely, the low frequency branch corresponds to the TK mode and has 
\begin{equation}
\omega_{-}^2 = \frac{\hbar \Omega}{4m} \frac{c^{2} k^{4}}{4\Omega^{2} + c^{2} k^{2}}.
\end{equation}
For large $k$, this reproduces the original TK frequency $\omega_{\rm TK} $, while for small $k$ it exhibits the quadratic behavior $\omega_{-} \simeq \sqrt{\hbar/16 m \Omega} c k^{2}$. The transition between $k^{2}$ and $k$ dependence occurs at $k \sim \Omega/c > R_{\perp}^{-1}$. This suggests that the effects of compressibility, characterizing the $k^{2}$ dependence, play a crucial role in the TK mode. Thus, this regime is distinguished from the usual incompressible TF regime as the ``soft'' TF regime. When the finite compressibility is included, the observed values of $\omega_{(1,0)}$ are well explained \citep{Baym2,Cozzini,Sonin}. First-principle simulations based on the GP formalism also agree excellently with the experimental data \citep{Mizushima2,Baksmaty}. 

In the LLL limit, corrections to the elastic shear modulus $C_{2}$ of the vortex lattice are important; with increasing $\Omega$, its value eventually reaches $C_{2}^{\rm LLL} \simeq (81/80\pi^{4}) m c^{2} n$ \citep{Baym2}. Using this value, Schweikhard {\it et al.} compared the measured $\omega_{(1,0)}$ with the theoretical prediction, finding a crossover of $\omega_{(1,0)}$ from the TF results to the LLL results, as shown in Fig. \ref{TKmode}(B). However, more detailed theoretical analysis in the LLL limit revealed that the value of the shear modulus is estimated as $0.1027 m n c^{2}$, which is a factor of 10 larger than $(81/80\pi^{4}) m n c^{2}$ \citep{Sinova,Sonin2,Cozzini2}. This indicates that even though the equilibrium properties in the experiment are consistent with the LLL picture, the data of the TK frequency are still far from the LLL limit. One possible explanation of this discrepancy is an underestimate of the rotation rate from the aspect ratio of the cloud due to the defocus of the imaging camera or the breakdown of the TF theory \citep{Watanabe3}.

\subsection{Vortices in an anharmonic potential}
For a rotating condensate with a frequency $\Omega$ in a harmonic potential $(1/2)m\omega_{\perp}^{2}r^{2}$, the centrifugal potential cancels the confinement, thus preventing a BEC from rotating at $\Omega$ beyond $\omega_{\perp}$. This restriction can be avoided by introducing an additional quartic potential, so that the combined trapping potential in the $xy$ plane becomes $V_{\rm ex}(r) = (1/2) m \omega_{\perp}^{2} ( r^{2}  + \lambda r^{4}/a_{\rm ho}^{2})$, where the dimensionless parameter $\lambda$ characterizes the relative strength of the quartic potential. The properties of a rotating condensate in an anharmonic potential have recently attracted a lot of theoretical attention \citep{Fetter,Lundh2,Fischer,Kasamatsu1,Kavoulakis1,Aftalion3,Jackson1,Jackson12,Fetter2,Danaila,Kim,Bargi,Fu}.

\begin{figure}[hbtp]
\begin{center}
\includegraphics[height=0.25\textheight]{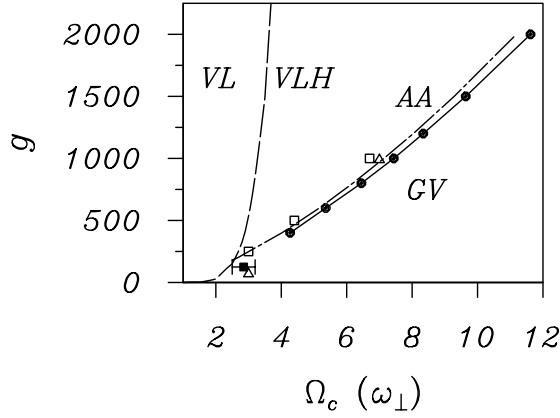}
\end{center}
\caption{Phase diagram of the vortex state with an additional quartic potential for $\lambda=1/2$ as a function of interparticle interaction strength ($g=4 \pi n_{z} a$) versus rotation rate ($\Omega_{c} = \Omega$). The dashed curve denotes the onset of a central density hole (VLH) in the uniform vortex lattice state (VL), obtained by TF analysis as $\Omega_{h}^{2} = 1+2 \sqrt{\lambda} (3 \sqrt{\lambda} g / 2 \pi)^{1/3}$ \citep{Fetter2}. The dashed-dotted curve and the solid points ($\bullet$) joined by solid lines show the phase boundary $\Omega_{c}$ between the annular condensate with a circular array of vortices (AA) and the giant vortex (GV) state, which are determined by two different analytical methods \citep{Fu}. The open triangles ($\triangle$) are the values of $\Omega_{c}$ determined by the GP solution \citep{Fetter2} and the open squares ($\square$) are the results using an improved variational approach \citep{Kim}. The filled square ($\blacksquare$) with error bars gives the approximate bounds on $\Omega_{c}$ determined numerically for $g=125$ \citep{Kasamatsu1}. For a weakly interacting limit $g \sim 10$, a much richer structure was revealed \citep{Jackson1,Jackson12}.  (Taken from \citep{Fu}. Reprinted with permission from APS.)}
\label{qurtphase}
\end{figure}
The vortex phases in an anharmonic trap are quite different from those in a harmonic trap, since it is possible to rotate the system arbitrarily fast. The predicted phase diagram of the vortex states as a function of interparticle interaction strength versus rotation rate is shown in Fig. \ref{qurtphase}. For small $\Omega$, the equilibrium state is the usual vortex lattice state. As $\Omega$ increases, the vortices begin to merge in the central region and the centrifugal force pushes the particles towards the edge of the trap. This results in a new vortex state consisting of a uniform lattice (multiple circular arrays of vortices) with a central density hole. The central hole becomes larger with increasing $\Omega$, and the condensate forms an annular structure with a single circular array of vortices. A further increase of $\Omega$ stabilizes a {\it giant vortex}, where all vortices are concentrated in the single hole \citep{Fischer}.

\begin{figure}[hbtp]
\begin{center}
\includegraphics[height=0.39\textheight]{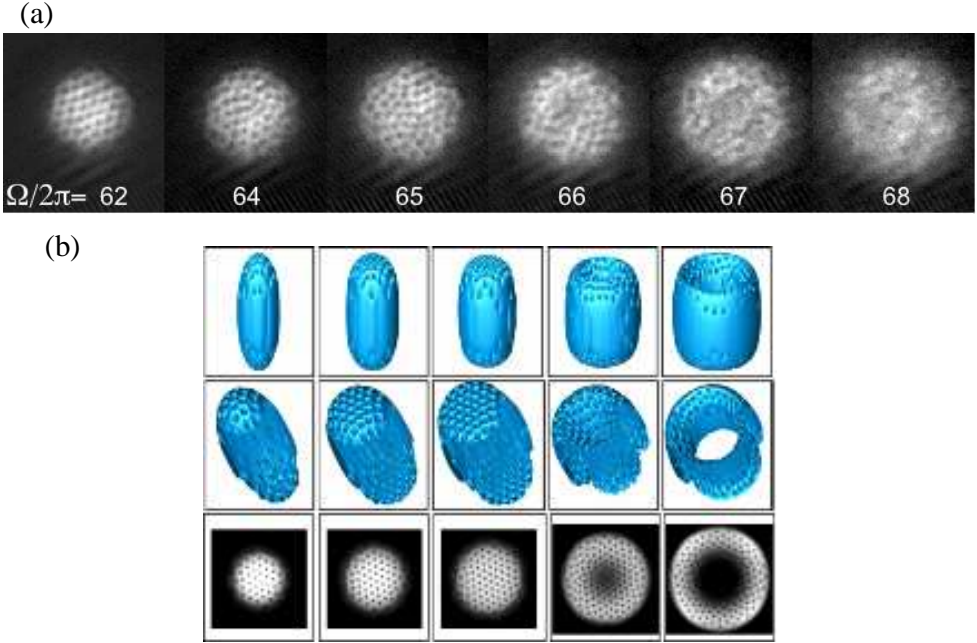}
\end{center}
\caption{(a) Density profiles of a rapidly rotating condensate in a quadratic plus quartic potential for various stirring frequencies $\Omega/2\pi$. For these data $\omega_\perp/2\pi=65$ Hz. (b) Ground state structure obtained by numerical simulations with a parameter corresponding to the experiment of Ref. \citep{Bretin2}. The rotation frequency is $\Omega / 2\pi =$ 60, 64, 66, 70.6, 73 Hz (respectively, $\Omega / \omega_{\perp} = $ 0.92, 0.98, 1.01, 1.08, 1.11) from left to right, where $\Omega / 2\pi = 70.6$ Hz corresponds to $\Omega_{h}$. The first two rows show 3D views of the vortex lattice as isosurfaces of low atomic density. In the bottom row, the density distribution is integrated along the $z$-axis. (Taken from \citep{Bretin2} and \citep{Danaila}. Reprinted with permission from APS.)}
\label{rotcrit}
\end{figure}
A combined harmonic-plus-quartic potential was formulated by the ENS group by superimposing a blue detuned laser with the Gaussian profile \citep{Bretin2}. Since the waist $w$ of the beam propagating along the $z$-axis is larger than the condensate radius, the potential created by the laser $U_{0} \exp(-2 r^{2}/w^{2})$ can be written as $U(r) \simeq U_{0} (1 - 2 r^{2} / w^{2} + 2  r^{4} / w^{4})$. The second term leads to a reduction of the transverse trapping frequency $\omega_\perp$ and the third term provides the desired quartic confinement, giving $\omega_{\perp}/2\pi=65$ Hz and $\lambda \simeq 10^{-3}$ for $V_{\rm ex}(r)$. Figure \ref{rotcrit} shows experimental images of the condensate density as the rotation frequency $\Omega$ is increased \citep{Bretin2}. For $\Omega <\omega_{\perp}$, the vortex lattice is clearly visible. When $\Omega >\omega_{\perp}$, however, the vortices becomes gradually difficult to observe and the images become less clear for $\Omega=1.05 \omega_{\perp}$ ($=2\pi \times 68$ Hz), which suggests a transition into a new vortex phase.

Despite the visibility of the cores, the angular momentum of the condensate monotonically increased, confirmed by the measurement of $R_{\rm TF}$ and surface wave spectroscopy. Hence, the most plausible explanation for this mysterious observation is that the vortex lines are still present, but strongly bent when $\Omega > \omega_\perp$. This bending may occur due to the finite temperature effect on the fragile vortex lattice at a high rotation rate; numerical simulations of the 3D GP equation show that, when looking for the ground state of the system using imaginary time evolution of the GP equation, much longer imaginary times were required to reach a well ordered vortex lattice for $\Omega >\omega_{\perp}$ than for $\Omega < \omega_{\perp}$. Compared with the numerical results shown in the bottom in Fig. \ref{rotcrit}, the condensate should still have an ordered visible lattice even for $\Omega \geq \omega_{\perp}$. To observe the density hole at the center, it is necessary to rotate at a slightly faster rate than the upper frequency used in this experiment.

\subsection{Vortex pinning in an optical lattice}
\begin{figure}[hbtp]
\begin{center}
\includegraphics[height=0.28\textheight]{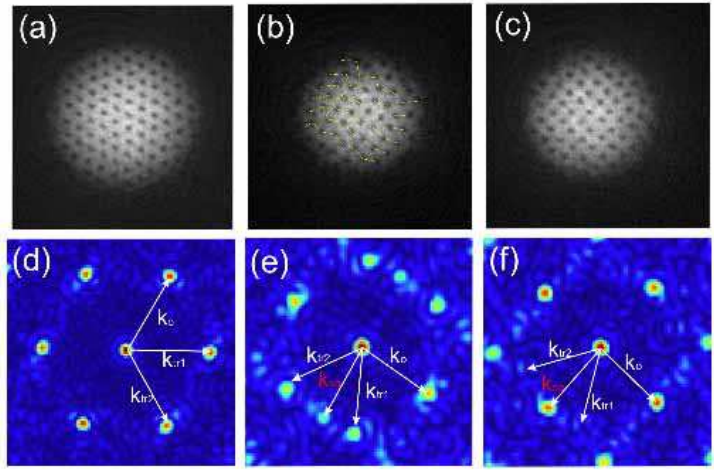}
\end{center}
\caption{Images of rotating condensates pinned to a co-rotating optical lattice at $\Omega=0.866 \omega_{\perp}$ with pinning strength $U_{\rm pin} / \mu = $ (a) 0.049, (b) 0.084, (c) 0.143, showing the structural crossover of the vortex lattice. (a)--(c) show absorption images of the vortex lattices after expansion. (d)--(f) are the Fourier transforms of the images in (a)--(c). $k_{0}$ is taken by convention to be the strongest peak; $k_{\rm tr1}$, $k_{\rm sq}$, and $k_{\rm tr2}$ are at 60$^\circ$, 90$^\circ$, and 120$^\circ$, respectively, from $k_{0}$. (Taken from \citep{Tung}. Reprinted with permission from APS.)}
\label{optlatpin}
\end{figure}

Rotating BECs combined with an optical lattice are an interesting system, which has two competing length scales, vortex separation and the periodicity of the optical lattice. The structure of the vortex lattice is strongly dependent on the externally applied optical lattice. Various vortex phases appear depending on the number of vortices per pinning center, i.e., the filling factor. The JILA group formulated a rotating optical lattice using a rotating mask \citep{Tung}. Such a rotating optical lattice provides a periodic pinning potential which is static in the corresponding rotating frame. The authors have observed a structural crossover from a triangular to a square lattice by increasing the potential amplitude of the optical lattice, as shown in Fig. \ref{optlatpin}. These observations are consistent with theoretical studies \citep{Reijnders,Pu2}. The rotating optical lattice provides new phenomena in vortex physics for rotating bosons; such as the realization of a driven vortex system in a periodic array \citep{Kasamatsu4} or strongly correlated phases in rotating bosons \citep{Bhat}.

\section{Other topics and future studies} \label{other}
In this section, we discuss other intriguing problems associated with quantized vortices in atomic BECs. Since many theoretical works have predicted novel properties of vortices under a variety of situations, it is impossible to refer to all studies in this section. We thus select some important issues that have been highlighted in the ultracold atom community or are likely to be investigated in experiments in the near future. 

\subsection{A vortex in an attractively interacting BEC}
Although we have dealt primarily with vortices in repulsively interacting BECs, vortices in attractively interacting BECs characterized by a negative s-wave scattering length $a<0$ have also received theoretical interest. A homogenous BEC with $a<0$ is never stable due to self-focusing collapse, but a BEC in a confining potential can remain stable as long as the number of condensed atoms $N$ lies below a critical value $N_{c} \sim a_{\rm ho}/|a|$ \citep{Bradley}. This is because the self-focusing can be balanced by the kinetic energy (quantum pressure), which tends to defocus the wave function. 

Since a central vortex state reduces the peak density, it may help stabilization of a trapped condensate with $a<0$ in the sense that it can contain a larger number of atoms by suppressing self-focusing. However, for $\Omega < \omega_{\perp}$ excitation of a vortex in a harmonically trapped BEC with $a<0$ is prohibited by the center-of-mass motion, which is the lowest energy state for a given angular momentum \citep{Wilkin}. Some authors have proposed that the use of an anharmonic confinement can support a stable vortex phase as well as regimes of center-of-mass motion \citep{Lundh3,Kavoulakis,Ghosh2,Collin}. The vortex state in this system can be regarded as a ring bright soliton with nonzero winding number \citep{Carr}, showing interesting splitting dynamics dominated by the dynamical instability of the quadrupole mode \citep{Saito}.

\subsection{Vortices in dipolar condensates}
BECs of chromium atoms have recently been created \citep{Griesmaier}, exhibiting a larger magnetic-dipole moment ($\mu_{d} = 6 \mu_{B}$; $\mu_{B}$ is the Bohr magneton) than those of typical alkali atoms ($\mu_{d} \simeq \mu_{B}$). This opens the door for studying the effect of anisotropic long-range interactions in BECs. The interaction potential between two magnetic dipoles $\mu_{d} \hat{\bf e}$ separated by ${\bf r}$ is given by $V_{dd}({\bf r})=(\mu_{0} \mu_{d}^{2}/4 \pi) (1 - 3 \cos^{2} \theta)/r^{3}$, where $\mu_{0}$ is the vacuum magnetic permeability and $\hat{\bf e} \cdot \hat{\bf r} = \cos \theta$. Such dipole--dipole interactions contribute to the GP equation as a nonlocal mean-field potential as $i \hbar \partial \psi /\partial t = [ -\hbar^2 \nabla^2 / 2 m +V_{\rm ex} + g |\psi|^{2} + \int d {\bf r}' V_{dd}({\bf r}'-{\bf r}) |\psi({\bf r}')|^{2}] \psi$. Since the scattering length can be tuned to zero by a Feshbach resonance technique, we can obtain novel quantum {\it ferrofluids} dominated by the dipole--dipole interaction \citep{Lahaye}.  

The principal effect of $V_{dd}$ on the equilibrium properties of a condensate is to cause distortion of its aspect ratio so that it is elongated along the direction of the dipoles. This feature affects the stability of a vortex in a dipolar BEC; the thermodynamic critical rotation frequency $\Omega_{c}$ decreases for a condensate in a pancake-shaped trap ($\omega_{\perp} < \omega_{z}$), while it increases a cigar-shaped trap ($\omega_{\perp} < \omega_{z}$), compared to that of a conventional BEC (Sec. \ref{vortexbasictheory}) \citep{ODell}. Interestingly, the critical frequency $\Omega_{c}$ can become larger than the onset of the dynamical instability of a rotating condensate (see Sec. \ref{formationdyn}). This is an intriguing regime where a rotating dipolar BEC is dynamically unstable but vortices will not enter \citep{Bijnen}. Numerical simulations show that the structure of vortices has a craterlike shape for $\hat{\bf e} \parallel \hat{\bf z}$ and has an elliptical shape for $\hat{\bf e} \perp \hat{\bf z}$ \citep{Pu}. Rapidly rotating dipolar BECs possess a rich variety of vortex phases characterized by different symmetries of the lattice structure \citep{Cooper,Zhang,Komineas2}.

\subsection{Melting state of vortex lattices: beyond the LLL regime}
At sufficiently high rotation rates, a vortex lattice should melt via quantum fluctuations \citep{Sinova} and the system should then begin to enter a strongly-correlated vortex liquid phase. Exact diagonalization studies for small number of bosons have revealed that the ground states exhibit strong analogy with the physics of electronic fractional quantum Hall states \citep{Viefers,Cooper2,Paredes,Regnault,Regnault2,Ghosh1}; for specific filling factors $\nu = N/N_{v}$, i.e., the ratio of the total number and the vortex number, the ground state possesses incompressibility characterized by the energy gap. For example, at angular momentum $L_{z} = N (N-1)$, where $N_v = 2N$, the exact ground state is an $N$-particle fully symmetric Laughlin wave function adopted for bosons: $\Psi({\bf r}_1, {\bf r}_2, \cdots, {\bf r}_N) \sim \prod_{j\ne k}(z_{j}-z_{k})^2 e^{-\Sigma_{l} r_l^2/2a_{\rm ho}^2}$ with $z_{j} = x_{j}+iy_{j}$. 

The conditions for the formation of these states have been expressed as $\nu \leq {\cal O}(1)$. It can be seen that an unrealistically high rotation is necessary to satisfy the condition; observed filling factors are always greater than 100 \citep{Schweikhard}, which are still deeply within the mean-field GP regime. To overcome this difficulty, insertion of a 1D optical lattice along the $z$-direction has been proposed to enhance the quantum fluctuations of the vortices \citep{Martikainen,Snoek}. Then, the optical lattice divides the condensate into pancake fractions coupled by a tunneling process between near neighbors and $N$ in a single pancake is greatly reduced. Achieving this regime experimentally remains an important challenge.

\subsection{Spontaneous vortex generation associated with phase transitions}
In 2D systems with continuous symmetry, true long-range order is destroyed by thermal fluctuations at any finite temperature. For 2D Bose systems, a quasi-condensate can be formed with a correlation decaying algebraically in space, where superfluidity is still expected below a certain critical temperature. This 2D phase transition is closely connected with the emergence of thermally activated vortex--antivortex pairs, known as the Berezinskii--Kosterlitz--Thouless (BKT) phase transition occurring at $T=T_{\rm BKT}$ \citep{Berezinskii,Kosterlitz}. For $T<T_{\rm BKT}$, isolated free vortices are absent; vortices always exist only in the form of bound pairs, formed by two vortices with opposite circulations. The contribution of these vortex pairs to the decay of the correlation is negligible, and the algebraic decay is dominated by phonons. For $T>T_{\rm BKT}$, the free vortices form a disordered gas of phase defects and give rise to an exponential decay of the correlation.

Recently, the BKT transition was observed experimentally in ultracold atomic gases \citep{Hadzibabic,Schweikhard3,Kruger,Hadzibabic2}. In the ENS experiment, a 1D optical lattice was applied to an elongated condensate, splitting the 3D condensate into an array of independent quasi-2D BECs \citep{Hadzibabic}. The interference technique revealed the temperature dependence of an exponent of the first-order correlation function of the fluctuating 2D bosonic field \citep{Polkovnikov}. A universal jump in the superfluid density characteristic of the BKT transition was identified by observing the sudden change of the exponent, where the finite size effect causes a finite-width crossover rather than a sharp transition. Surprisingly, the microscopic origin of this transition, i.e., whether or not it is a BKT type transition, was directly clarified from the image of the interference of the two 2D condensates. If isolated free vortices are present in either of two condensates, the interference fringes exhibit dislocations. Such a dislocation has been observed in the high-$T$ region of the crossover, supported by the theory using classical field simulations \citep{Simula3}. 

In contrast, the JILA group inserted a 2D optical lattice into a condensate to create 2D bosonic Josephson junction arrays \citep{Schweikhard3}. Each condensate was localized at a site $j$. Each had an individual phase $\theta_{j}$ and was separated by a potential barrier from the nearest neighbors. This system can be mapped to the XY model, $H=-J \sum_{\langle j,j' \rangle} \cos (\theta_{j} - \theta_{j'})$, where $J$ denotes the tunneling coupling and the sum is restricted within nearest neighbors \citep{Trombettoni}. The XY model is expected to exhibit a BKT transition at $T_{\rm BKT} \simeq J$ from free-energy considerations. Direct imaging of vortex cores and the systematic determination of $J$ revealed evidence for a gradual increase in the number of isolated free vortices at $T \geq J$, consistent with the BKT crossover picture. 

A related work is vortex formation by merging three uncorrelated BECs that are initially separated by a triple well potential \citep{Scherer}. Depending on the relative phases between the condensates and merging rate, vortices formed stochastically without applying rotation. This situation is useful to clarify the mechanism of spontaneous vortex generation through the Kibble-Zurek mechanism \citep{Kibble,Zurek} during rapid phase transition \citep{Leggett2,Kasamatsu0}. 

\subsection{Skyrmions in multi-component BECs}
Another important issue in vortex physics is to elucidate the vortex phases in multicomponent (spinor) BECs. Multicomponent order parameters allow the formation of various unconventional topological defects with complex properties that arise from interactions between different order parameter components. Since it is possible to load and cool atoms in more than one hyperfine spin state or more than one atomic element in the same trap, multicomponent condensates can be realized experimentally. Such systems offer an ideal testing ground for the study of unconventional topological defects; similar structures appear in other condensed matter systems such as superfluid $^{3}$He and unconventional superconductors, and theories in high-energy physics and cosmology. A few experimental works have investigated the properties of composite vortices in spinor BECs \citep{Leanhardt2,Schweikhard2,Sadler}. A review of this topic is presented in Ref. \citep{Kasamatsurev} and references therein.

\subsection{Vortices in Fermion condensates}
Quantum degenerate Fermi gases provide a remarkable opportunity to study strongly interacting fermions. In contrast to other Fermi systems, such as superconductors, neutron stars, or the quark--gluon plasma, these gases have low densities and their interactions can be precisely controlled over a wide range by using a Feshbach resonance technique. For small and negative values of the scattering length $a$ the equation of state approaches the limit of a noninteracting Fermion gas, while for small and positive values the system behaves as bosons of tightly-bound molecules. Therefore, we can study the crossover from a BEC of molecules to a Bardeen--Cooper--Schrieffer (BCS) superfluid of loosely-bound Cooper pairs when an external magnetic field is varied across a Feshbach resonance. Recent topics in this rapidly growing field are presented in the comprehensive review paper Ref. \citep{Giorgini}. 

Decisive evidence for fermion superfluidity was obtained from observations of long-lived vortex lattices in a strongly-interacting rotating Fermion gas \citep{Zwierlein}. Rotation was applied to an ultracold gas of $^{6}$Li atoms in $| F=1/2, m_{F} = \pm 1/2 \rangle$, in a similar way as for conventional BECs (see Sec. \ref{stirringexp}), with magnetic fields covering the entire BEC--BCS crossover region. A crucial problem was detecting the vortex cores in the BCS limit, because a sufficient density depletion at the vortex core could not be expected \citep{Nygaard}. In the experiment, the visibility of the vortex cores was increased by a rapid sweep of the magnetic field from the BCS to the BEC side during ballistic expansion of the TOF measurement. These measurements strongly support the existence of vortices before the expansion even on the BCS side of the resonance. 

The most striking aspect of this experiment is that it opens up the possibility of studying vortex physics in a strongly-coupled fermion superfluid in a systematically controlled way. In the strong-coupling limit $|a| \rightarrow \infty$ at the resonance, called a unitarity limit, the Fermi gas exhibits universal behavior. Along this line, several microscopic calculations of the vortex structure, based on the BdG formalism, have been carried out \citep{Bulgac,Sensarma,Chien,Machida,Machida2}. More detailed studies may provide useful predictions of the mysterious vortex properties in high-$T_{c}$ superconductors, and eventually those in room-temperature superconductors. 

\section{Conclusion} \label{conclusions}
Quantized vortices in atomic Bose--Einstein condensates constitute an active research field, which has drawn the continuous attention of researchers in related fields such as superconductors, mesoscopic systems, nonlinear optics, atomic nuclei, and cosmology, as well as superfluid helium. Quantized vortices in rotating condensates have provided conclusive evidence for superfluidity because they are a direct consequence of the existence of a macroscopic wave function that describes the superfluid. The direct imaging of the vortex cores and lines helps us understand the fundamentals of superfluid dynamics. Especially, the inhomogeneous effect caused by a confining potential yields new features in both a slowly rotating regime and a rapidly rotating one, not found in a bulk superfluid system. The observed phenomena are consistent with the prediction of the Gross--Pitaevskii equation without fitting parameters. 

Finally, this volume addresses mainly the topics of {\it quantum turbulence}. The feasibility of generating quantum turbulence in a trapped BEC is discussed by one of the authors \citep{Kobayashi}. 
\\

Acknowledgments

K.K. acknowledges the support of a Grant-in-Aid for Scientific Research from JSPS (Grant No. 18740213). M.T. acknowledges the support of a Grant-in-Aid for Scientific Research from JSPS (Grant No. 18340109) and a Grant-in-Aid for Scientific Research on Priority Areas (Grant No. 17071008) from MEXT.

% \appendix
% \pagestyle{myheadings}

% \section{}
% \label{}

% Bibliographic references with the natbib package:
% Parenthetical: \citep{Bai92} produces (Bailyn 1992).
% Textual: \citet{Bai95} produces Bailyn et al. (1995).
% An affix and part of a reference:
%   \citep[e.g.][Ch. 2]{Bar76}
%   produces (e.g. Barnes et al. 1976, Ch. 2).

\end{document}